\documentclass[%
 reprint,
 amsmath,amssymb,
 aps,
 pra,
 superscriptaddress]{revtex4-2}

\usepackage[T1]{fontenc}
\usepackage[colorlinks=true, allcolors=blue]{hyperref}
\usepackage{amsmath}
\usepackage{braket}
\usepackage{graphicx}
\usepackage{url}
\usepackage{dcolumn}
\usepackage{bm}
\usepackage{xcolor}

\newcommand{\tr}{\mathrm{tr}}

\begin{document}

\title{Bubble-Sort Exchange Networks: Symmetry Selection Rules and Language-Model Distillation on Exchange-Only Spin Chains}

\author{Yuichiro Minato}
\email{minato@blueqat.co.jp}
\affiliation{blueqat Inc., 2-24-12 Shibuya, Shibuya-ku, Tokyo 150-6139, Japan}

\begin{abstract}
Linear arrays of exchange-coupled quantum dots have a single fast two-body control, the Heisenberg exchange pulse, and a pulse of area $\pi$ \emph{is} a SWAP. The parallel bubble-sort SWAP network that routes all-to-all interactions on a superconducting chain is therefore native to exchange-only (EO) spin qubits, and a pulse of any area that is not a multiple of $\pi$ fuses the routing with an entangling interaction. We build a variational ``bubble-sort exchange network'' on a chain of $N$ dots---$R$ rounds of odd--even transposition, one real parameter per pulse---and study it as a Born machine read out by Pauli spin blockade (PSB). Exchange conserves total spin, so what the readout can reach is fixed by the total-spin content of the initial product state. For the two-spin singlet--triplet encoding at zero magnetisation we derive a selection rule---the probability of outcomes with $k$ triplet pairs is bounded by the initial weight in sectors $S\le k$ ($S=0$ excluded when $k=1$), and likewise for any set of classes by the union of their sectors---and prove by a transportation argument that it is also sufficient: the closure of the reachable readout distributions is the polytope cut out by the rule, and the implied total-variation floor is attained. Products of singlets and $T_0$'s forbid whole outcome classes; the N\'eel state $\ket{\uparrow\downarrow\uparrow\downarrow\cdots}$ forbids none, and the three-spin EO encoding has no floor from a code state. We then distil, as a proof of concept and in noise-free simulation, short next-token distributions of GPT-2 over restricted vocabularies into the network, with matrix-product-state (MPS) Born machines as the classical control. At six logical qubits the three-spin chain reaches $\mathrm{KL}=0.011$~nat at depth $N$ and $0.004$ at depth $2N$, the latter below the bond-dimension-4 MPS ($0.005$); the two-spin chain from the N\'eel state reaches $0.014$ at depth $2N$ and $0.004$ at depth $3N$ (medians of five seeds), against $0.28$--$1.5$ from products of singlets and $T_0$'s for the first teacher (for the second, whose forbidden class carries negligible mass, the alternating product state is competitive). Final states lie largely outside the code space: for a Born machine the leakage space is a resource, not an error. No arrangement of disjoint pulses, in any geometry, brings every pair of spins together in fewer than $N-1$ rounds, one fewer than the chain's $N$, and at depth $2N$ the chain distils as well as or better than a two-dimensional array or non-local couplings in our tests. Sector-restricted simulation with an adjoint-method gradient extends the picture to ten to fourteen logical qubits (1024--16384 outcomes, up to 30 dots): at fixed depth in units of $N$ the two-spin chain loses ground to the MPS as the size grows (for the first teacher, from the $\chi=4$ level at six qubits to the $\chi=3$ level at twelve), while the three-spin chain again reaches the $\chi=4$ level at depth $2N$ for the first teacher. As a side result, the orthogonal factor of the polar decomposition of a two-qubit weight block compiles into the same network at depth $4N$ (Appendix~\ref{sec:unitary}).
\end{abstract}

\maketitle

\section{Introduction}

Linear arrays of exchange-coupled quantum dots are the simplest scalable geometry for semiconductor spin qubits, and Heisenberg exchange is their only fast, all-electrical two-body control. Two encodings exploit this: the three-spin exchange-only (EO) qubit \cite{DiVincenzo2000,Bacon2000,Weinstein2023}, which is universal under exchange alone at the price of a large leakage space, and the two-spin singlet--triplet (S--T) qubit \cite{Petta2005,Levy2002}, which is more compact but needs a magnetic gradient for universal single-qubit control. In both cases the elementary operation is the same: $U_{ij}(\theta)=\exp(-i\theta\,\mathbf S_i\!\cdot\!\mathbf S_j)$ on neighbouring dots.

A separate line of work in superconducting hardware addresses a different problem, connectivity. To run an all-to-all Ising cost function on a linear chain, Yan \emph{et al.} \cite{Yan2022} (following Harrigan \emph{et al.} \cite{Harrigan2021} and Hirata \emph{et al.} \cite{Hirata2009}) route the interactions with a parallel bubble sort: $n$ rounds of odd--even transposition, each round applying a ZZ$\cdot$SWAP block to every disjoint neighbouring pair, so that after $n$ rounds every pair of logical qubits has been adjacent exactly once and the register is reversed. Each ZZ$\cdot$SWAP block is a ZZ module (two CNOTs around an $R_z$) followed by a SWAP, which Ref.~\cite{Yan2022} counts as a depth-4 circuit before native compilation, i.e.\ depth $4n$ for the sort.

\begin{figure*}[t]
\includegraphics[width=\textwidth]{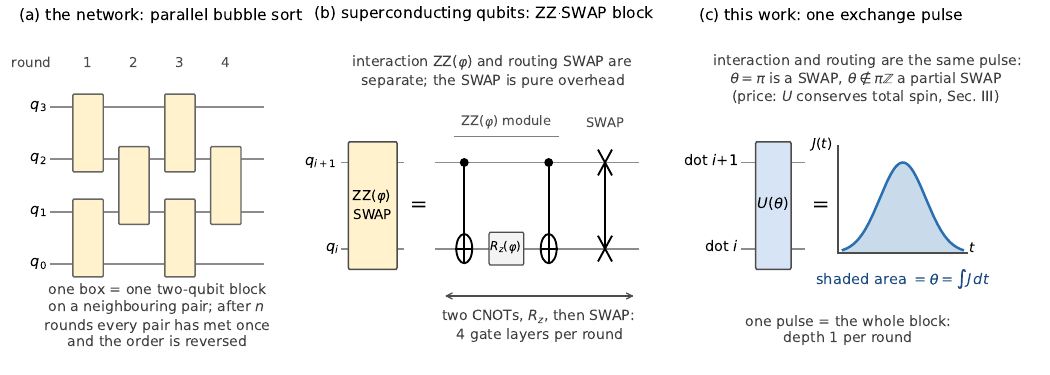}
\caption{Where this paper starts. (a) The parallel bubble sort (odd--even transposition): in each round a two-qubit block acts on every other neighbouring pair, the pairs alternate from round to round, and after $n$ rounds every pair of qubits has been adjacent exactly once while the register order is reversed. (b) On superconducting hardware \cite{Yan2022} the block is ZZ$(\varphi)\cdot$SWAP: the interaction (two CNOTs around an $R_z$) and the routing SWAP are separate operations, four gate layers per round in the count of Ref.~\cite{Yan2022}, before native compilation. (c) On an exchange-coupled spin chain the block is a single exchange pulse $U(\theta)=e^{-i\theta\,\mathbf S_i\cdot\mathbf S_{i+1}}$ whose only parameter is the pulse area $\theta=\int J\,dt$: $\theta=\pi$ is a SWAP, any $\theta\notin\pi\mathbb Z$ a partial SWAP that routes and entangles at once, so one round has depth 1 and the whole network is a parametrised circuit with one number per pulse. The price is that $U(\theta)$ conserves total spin (Sec.~\ref{sec:rule}).}
\label{fig:zzswap}
\end{figure*}

The observation behind this paper is that the same network is free on an exchange-coupled chain (Fig.~\ref{fig:zzswap}). Since $\mathbf S_i\!\cdot\!\mathbf S_j=\tfrac12(\mathrm{SWAP}_{ij}-\tfrac12)$,
\begin{equation}
U_{ij}(\theta)=e^{i\theta/4}\bigl[\cos\tfrac{\theta}{2}\,\mathbb 1-i\sin\tfrac{\theta}{2}\,\mathrm{SWAP}_{ij}\bigr],
\label{eq:exchange}
\end{equation}
so $\theta=\pi$ is a SWAP up to a phase, $\theta=0$ is the identity, and every intermediate $\theta$ is a partial SWAP that both entangles and routes: writing $U(\theta)=e^{-i\pi/4}\,\mathrm{SWAP}\,U(\theta-\pi)$, each pulse is a SWAP followed by a residual exchange of area $\theta-\pi$ acting on the routed labels. A bubble-sort SWAP network is therefore not an overhead to be paid for connectivity but a parametrised circuit family in its own right: $R$ rounds of odd--even transposition with one pulse area per pair, which at $\theta\equiv\pi$ and $R=N$ reduces to the reversal permutation and at $\theta\equiv0$ to the identity. We call this family a \emph{bubble-sort exchange network}. It has depth $R$ and $\approx R(N-1)/2$ pulses on $N$ spins, and at $R=N$ every pair of spins has interacted exactly once.

The idea has neighbours in three literatures. Fermionic swap networks \cite{Kivlichan2018,OGorman2019,Hashim2022} and the ``efficient Hamiltonian variational ansatz'' of Cade \emph{et al.} \cite{Cade2020} use the swap-network layout as the scaffold of a variational circuit, but keep the swap itself parameter-free; the Hartree--Fock experiment of Arute \emph{et al.} \cite{Arute2020} trains a linear-depth network of Givens rotations, i.e.\ partial fermionic swaps, which is classically simulable as a matchgate circuit, and swap networks have recently been embedded into layered ans\"atze for arbitrary connectivity \cite{Parella2025}. On the spin-qubit side, Kattem\"olle and van Wezel \cite{Kattemolle2022} note that the Heisenberg gate is native to quantum dots and that the SWAP is an exchange pulse of area $\pi$ up to a phase, and use fixed SWAPs to route a Heisenberg variational eigensolver on a kagome lattice, while Zheng \emph{et al.} \cite{Zheng2023} build permutation-equivariant learning ans\"atze from exponentials of transpositions, and general SU(2)-equivariant circuit constructions are given in Refs.~\cite{Meyer2023,East2026}. Closest in spirit, Burkat \emph{et al.} \cite{Burkat2026} recently formalised circuits made of computational-basis inputs, arbitrary-angle exchange gates and computational-basis readout as the class XQP, showed that $\sqrt{\rm SWAP}$ circuits are semi-universal within the SU(2) commutant, and proved that additive-error classical simulation of such circuits would imply efficient simulation of BQP; the bubble-sort exchange network is an XQP circuit with a fixed layout, learned angles and spin-blockade readout. Exchange-driven SWAP chains have been demonstrated on quantum-dot arrays \cite{Kandel2019,Qiao2020,vanDiepen2021}, long SWAP sequences on 20--100-dot chains have been simulated with tensor networks \cite{Taylor2024}, and resource estimates for standard algorithms on exchange-only hardware have recently been given \cite{Lohof2026}. What is new here is to make the swap angle itself the parameter on a linear exchange chain, to use the resulting SU(2)-symmetric network as a Born machine with spin-blockade readout, and to work out what that symmetry allows.

Our interest is machine learning rather than optimisation \cite{Minato2024,Sotobayashi2025}, and specifically \emph{distillation}: taking a probability distribution produced by a classical model---here the distribution of two- or three-token continuations of GPT-2 over a restricted vocabulary---and re-learning it in a Born machine whose samples are produced by projective readout of the spin chain. This is the setting of our earlier work on TE-EX MNIST \cite{Minato2026}, and it connects to the tensor-network route into quantum circuits: a classical model or its output distribution is first compressed into a matrix-product state (MPS) \cite{Han2018,Ran2020,Rudolph2023,Dborin2022}, which is then used to initialise or to guide a parametrised circuit. On an exchange chain, however, one has to ask first what can be reached at all. Exchange commutes with total spin, and the natural readout, Pauli spin blockade (PSB) on a dot pair \cite{Petta2005}, distinguishes singlet from triplet; both are SU(2)-structured, and their interplay constrains the achievable readout distributions independently of circuit depth. The first half of the paper works this out; the second half uses it to design the distillation.

Our contributions are:
\begin{enumerate}
\item A closed-form \emph{selection rule} for PSB readout under exchange-only dynamics (Sec.~\ref{sec:rule}): for any set $K$ of Hamming classes of the outcome string (number of triplet pairs), $P(|b|\in K)\le\sum_{S\in\Lambda(K)}c_S^2$, where $c_S^2$ is the weight of the initial state in total-spin sector $S$ and $\Lambda(k)=\{0,\dots,k\}$ for $k\ne1$, $\Lambda(1)=\{1\}$. The rule is also sufficient (Sec.~\ref{sec:suff}): since the readout projectors commute with total spin, the reachable distributions are the marginals of a transportation problem from sectors to outcomes, and max-flow/min-cut turns Eq.~(\ref{eq:rule}) into an exact characterisation of the closure of the reachable set, with the implied total-variation (TV) floor attained. A direct numerical search over the symmetry-allowed states agrees to four digits in all ten cases tested (Sec.~\ref{sec:tight}).
\item The consequence for the two encodings: the three-spin EO logical space lies in a single sector and has no floor; the two-spin S--T encoding with any product of singlets and $T_0$'s has a parity-restricted sector content and hence structural zeros, but the N\'eel product state $\ket{\uparrow\downarrow\uparrow\downarrow\cdots}=\bigotimes(\ket S+\ket{T_0})/\sqrt2$ contains every sector, leaves no outcome class forbidden, and has a negligible floor for the teachers studied.
\item End-to-end distillation of GPT-2 distributions (Sec.~\ref{sec:distil}): six three-spin EO qubits at depth $R=N=18$ reach $\mathrm{KL}=0.011$--$0.014$ nat (within a factor of three of the bond-dimension-4 MPS, $0.005$; $\chi=1$: $0.35$, $\chi=2$: $0.14$), and six two-spin S--T qubits reach $0.014$ at depth 24 and $0.004$ at depth 36 (medians of five seeds; best seed $0.003$) with the N\'eel state versus $0.28$--$1.5$ from products of singlets and $T_0$'s for the first teacher, with final states $55$--$99\%$ outside the code space---the ``leakage shortcut''. Sector-restricted simulation extends this to 24-dot chains (Sec.~\ref{sec:scale}), and an adjoint-gradient GPU simulator to 30-dot chains with 10--14 logical qubits (Sec.~\ref{sec:e10}), where the three-spin chain still reaches the $\chi=4$ level at depth $2N$ for the first teacher.
\item A depth bound and its numerical counterpart (Sec.~\ref{sec:connect}): all-to-all interaction needs $N-1$ rounds of disjoint pulses in any geometry, and at depth $2N$ the chain distils as well as or better than a two-dimensional array, a non-local round robin of all pairs, and random matchings---at eight qubits by a factor $1.2$--$1.6$ in the median over five seeds---while at depth $N$ it is within $20\%$ of the best of them.
\item As a side result (Appendix~\ref{sec:unitary}): leakage-free preparation of the twelve two-qubit logical states tested at $R=N+2$ ($F=1-10^{-8}$) and compilation of the four two-qubit logical unitaries tested---three random polar factors and the polar factor $Q$ of a GPT-2 weight block---at $R=4N$ ($F_{\rm proc}=1-10^{-6}$); three-qubit unitaries were not reached at that depth by our optimisation.
\end{enumerate}

\section{Bubble-sort exchange networks}
\label{sec:network}

\subsection{Chain, pulses, schedule}
We consider $N$ spin-1/2 particles on a line with the Heisenberg Hamiltonian $H(t)=\sum_i J_i(t)\,\mathbf S_i\!\cdot\!\mathbf S_{i+1}$ and piecewise-constant pulses, so that each pulse on pair $(i,i{+}1)$ implements $U_{i,i+1}(\theta)$ of Eq.~(\ref{eq:exchange}) with area $\theta=\int J\,dt$. Pulses on disjoint pairs commute and are applied in parallel. The \emph{bubble-sort schedule} with $R$ rounds is
\begin{equation}
\mathcal B_R=\prod_{r=0}^{R-1}\ \prod_{i\equiv r\ (\mathrm{mod}\ 2)}U_{i,i+1}(\theta_{r,i}),
\end{equation}
i.e.\ round $r$ acts on the pairs $(0,1),(2,3),\dots$ for even $r$ and $(1,2),(3,4),\dots$ for odd $r$. Every pulse carries its own parameter. For even $N$ the network has $R\cdot(N-1)/2$ parameters when $R$ is even ($N/2$ and $N/2-1$ pulses in alternate rounds); for odd $N$ every round has $(N-1)/2$ pulses and the count is $R(N-1)/2$ for any $R$.

\begin{figure}[t]
\includegraphics[width=\columnwidth]{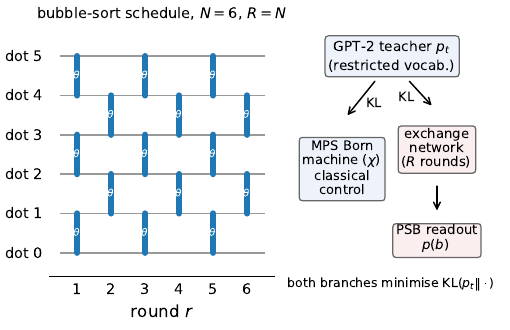}
\caption{Left: the bubble-sort (odd--even transposition) schedule for $N=6$ dots, $R=N$ rounds. Each bar is one exchange pulse $U(\theta)$ with its own pulse area; $\theta=\pi$ on every bar reverses the chain. Right: the distillation pipeline used in Sec.~\ref{sec:distil}: a restricted-vocabulary GPT-2 continuation distribution is fitted independently by MPS Born machines (a classical control for the bond dimension the circuit must emulate) and by the exchange network, trained directly on the KL divergence to the teacher, read out by Pauli spin blockade on the first pair of every block.}
\label{fig:schedule}
\end{figure}

\subsection{Elementary properties}
\begin{itemize}
\item[(i)] \emph{Reversal and exact cover.} At $\theta_{r,i}\equiv\pi$, $\mathcal B_N$ is the reversal permutation $i\mapsto N-1-i$ up to a global phase, and every pair of particles is swapped exactly once, so the $N(N-1)/2$ pulses are in bijection with the edges of the complete graph $K_N$ \cite{Yan2022,Hirata2009}. We verified this numerically for $N=6,8,12$; $R=N-1$ rounds do not suffice.
\item[(ii)] \emph{Identity initialisations.} $\theta\equiv0$ is the identity and $\theta\equiv\pi$ is a permutation; for $R$ a multiple of $N$ the latter maps blocks onto blocks, reversed internally, which maps each code space onto itself (as a sign for the pair code, as a logical rotation for the three-spin code), so both give an exactly leakage-free, non-random starting point for optimisation. In the distillation runs below we compare random, $0$ and $\pi$ initialisations.
\item[(iii)] \emph{Depth.} One round is one pulse layer; this is an ideal logical depth, which assumes that disjoint pulses of different areas can be applied simultaneously within one layer time. The corresponding superconducting ZZ$\cdot$SWAP round of Ref.~\cite{Yan2022} has depth 4, so an $N$-round network has depth $N$ on the exchange chain versus $4N$ there. Unlike ZZ$\cdot$SWAP combined with single-qubit rotations, however, exchange pulses alone cannot generate arbitrary two-qubit unitaries: $U(\theta)$ is SU(2)-invariant. This is the price, and the subject of the next section.
\item[(iv)] \emph{Symmetry.} Every $U_{ij}(\theta)$ commutes with the total spin $\mathbf S^2$ and $S_z$, hence so does $\mathcal B_R$.
\item[(v)] \emph{Depth bound.} A spin takes part in at most one pulse per round, so under \emph{any} schedule of disjoint pulses---on a chain, on a two-dimensional array, or between arbitrary pairs---at least $N-1$ rounds pass before every pair of spins has interacted. For odd $N$ the pulse count sharpens the bound to $N$ rounds. The odd--even transposition on the chain needs $N$ rounds, one more than the bound for even $N$ (a non-local round robin attains $N-1$ there). The bound concerns direct pairwise encounters only, not the depth needed for any particular computation; whether a two-dimensional layout or a non-local coupling optimises better at a given depth is tested in Sec.~\ref{sec:connect}.
\end{itemize}

We also use a \emph{staircase} schedule (layers of $U_{0,1},U_{1,2},\dots,U_{N-2,N-1}$ applied sequentially, $L$ layers) as a control; it has $L(N-1)$ parameters but depth $L(N-1)$.

\subsection{Encodings and readout}
\label{sec:enc}
\begin{figure*}[t]
\includegraphics[width=\textwidth]{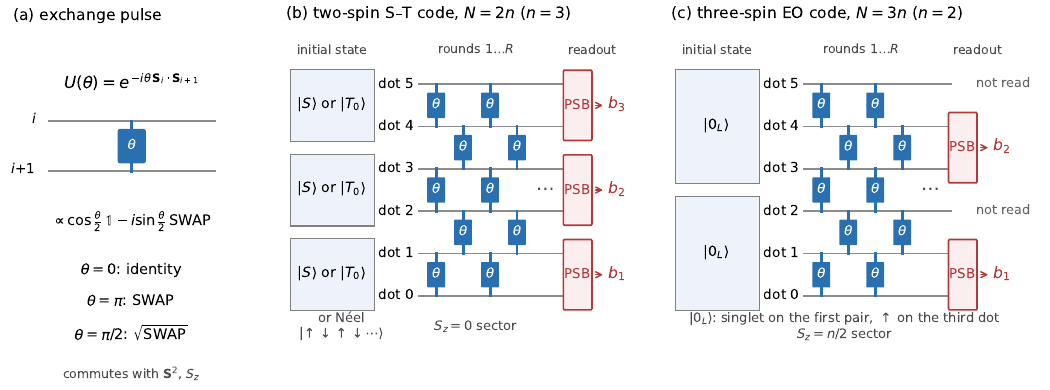}
\caption{The network as a circuit. (a) The only gate: one exchange pulse of area $\theta$ on a neighbouring pair, Eq.~(\ref{eq:exchange}); $\theta=0$, $\pi/2$ and $\pi$ give the identity, $\sqrt{\rm SWAP}$ and SWAP. (b) Two-spin S--T encoding: $N=2n$ dots in blocks of two, each block initialised in $\ket S$ or $\ket{T_0}$ (or the dots in the N\'eel product state), $R$ bubble-sort rounds of pulses on alternating pairs, and Pauli-spin-blockade readout of each block pair, giving the bit string $b=(b_1,\dots,b_n)$. (c) Three-spin EO encoding: blocks of three dots initialised in the logical state $\ket{0_L}$; the readout acts on the first pair of each block and the third dot is not read. In the distillation experiments the pulses are the only trainable parameters, leakage out of the code spaces is never penalised, and only $p(b)$ is compared with the teacher.}
\label{fig:circuit}
\end{figure*}

Two encodings are compared throughout, both on blocks of neighbouring dots (Fig.~\ref{fig:circuit}):

\emph{Two-spin S--T} (\texttt{st}): $\ket{0_L}=\ket S=(\ket{\uparrow\downarrow}-\ket{\downarrow\uparrow})/\sqrt2$, $\ket{1_L}=\ket{T_0}$, $N=2n$ spins for $n$ logical qubits, $S_z=0$.

\emph{Three-spin EO} (\texttt{eo}) \cite{DiVincenzo2000}: $\ket{0_L}=\ket S_{12}\ket\uparrow_3$, $\ket{1_L}=(2\ket{\uparrow\uparrow\downarrow}-\ket{\uparrow\downarrow\uparrow}-\ket{\downarrow\uparrow\uparrow})/\sqrt6$, the $S=\tfrac12$, $m=+\tfrac12$ doublet of three spins; $N=3n$, $S_z=n/2$.

\emph{Readout} is Pauli spin blockade on the first pair of every block: outcome $b_j=0$ if the pair is a singlet, $b_j=1$ if it is in any triplet state. PSB is a projective measurement of the pair-singlet projector $\Pi_{\rm s}$ versus $\mathbb 1-\Pi_{\rm s}$ (an idealisation of the transport measurement) and is well defined on any state, leaked or not; the outcome distribution $p(b)=\|\Pi_b\psi\|^2$ with $\Pi_b=\bigotimes_j\Pi^{(b_j)}_j$, $\Pi^{(0)}=\Pi_{\rm s}$, $\Pi^{(1)}=\mathbb 1-\Pi_{\rm s}$, is what the Born machine samples; $\mathrm{TV}(p,q)=\tfrac12\sum_b|p(b)-q(b)|$ throughout. On code states it coincides with the logical computational-basis measurement for both encodings.

\emph{Leakage} is $1-\|P_{\rm code}\psi\|^2$. We never penalise it in the distillation experiments: the leakage space is used freely, and only the readout distribution matters. For state preparation and unitary compilation (Appendix~\ref{sec:unitary}) leakage of the final state is what the fidelity measures.

\section{Symmetry ceilings}
\label{sec:ceiling}

\subsection{State preparation}
Let $P_S$ project onto total spin $S$ within the fixed $S_z$ sector. Since $\mathcal B_R$ is block-diagonal in $S$, the reachable unitaries are contained in $\bigoplus_S U(m_S)$, where $m_S$ is the multiplicity of sector $S$ at the given $S_z$; exchange generates the full commutant of SU(2) on each multiplicity space \cite{Kempe2001,Zheng2023,Hulse2024} (the corresponding random circuits form unitary designs of an order fixed by locality \cite{Mitsuhashi2025}), but not all relative phases between sectors \cite{Marvian2022,Marvian2024,Kazi2024,Burkat2026}. Readout distributions are blind to those phases (Sec.~\ref{sec:suff}); state fidelities are not, so the bound below is an upper bound whose attainment depends on the phases a given circuit can produce. For any target $t$ and initial state $\psi_0$
\begin{equation}
\max_{\mathcal B}\ |\braket{t|\mathcal B|\psi_0}|^2\ \le\ \Bigl(\sum_S\|P_St\|\,\|P_S\psi_0\|\Bigr)^2\equiv F_{\max}.
\label{eq:ceiling}
\end{equation}
For the three-spin encoding the whole logical space sits inside the single sector $S=n/2$ (every logical basis state is a product of $m=+\tfrac12$ block-doublet states, i.e.\ of highest-weight vectors, and therefore has $S=n/2$ at $S_z=n/2$), so $F_{\max}=1$ for every logical target from $\ket{0_L\cdots0_L}$. For the two-spin encoding the logical space is spread over $S=0,\dots,n$ with weights depending on the target, and a product initial state of singlets and $T_0$'s has a definite parity of $S$ (below), so $F_{\max}$ is typically $0.3$--$0.5$ for Haar-random logical targets at $n=3$. Table~\ref{tab:reach} shows that optimised networks approach these ceilings in the cases tested: for \texttt{st}, $R=2N$ saturates $F_{\max}$ to four digits (so the phases needed there are available); for \texttt{eo} with $n=2$ ($N=6$), $R=N+2=8$ rounds (20 pulses) prepare every Haar-random or MPS logical target tested to $F=1-10^{-8}$ (leakage $10^{-8}$), while $R=6,7$ do not, and the staircase needs $L=4$ ($20$ pulses) for the same. The multiplicity space of the $S=1$, $S_z=1$ sector of six spins is 9-dimensional, so 16 real parameters are needed; $R=7$ (18 pulses) falls short in practice, a reminder that the parameter count is necessary but not sufficient. For $n=3$ ($N=9$, 48-dimensional $S=\tfrac32$ multiplicity space, 94 real parameters needed for an arbitrary state in it) $R=9,14,18,22,27,32$ ($36,56,72,88,108,128$ pulses) give $F=0.923,\,0.995,\,0.9999,\,0.9999,\,1-10^{-5},\,1-10^{-10}$ (leakage $2\times10^{-2}$, $3\times10^{-3}$, $8\times10^{-5}$, $2\times10^{-4}$, $10^{-5}$, $10^{-10}$): below the parameter count ($R\le22$) the logical targets are approached to $10^{-4}$ and preparation to $10^{-5}$ ($R=27=3N$) and $10^{-10}$ ($R=32$) was observed only once the count is exceeded; the count is a heuristic, neither necessary (the logical targets form a 14-dimensional family) nor sufficient.

\begin{figure}[t]
\includegraphics[width=\columnwidth]{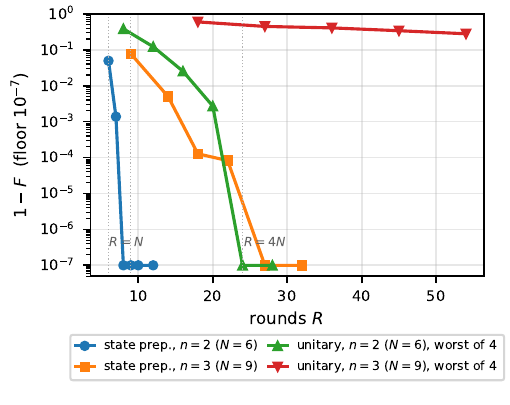}
\caption{Infidelity of the three-spin network versus number of rounds. State preparation: mean over Haar-random logical targets for $n=2$ ($N=6$) and $n=3$ ($N=9$); values below $10^{-7}$ are shown at the floor. Unitary compilation: worst of the four targets of Appendix~\ref{sec:unitary} (three random polar factors and the GPT-2 block) for $n=2$ and $n=3$. At $n=2$ a state needs $R\approx N+2$ and a leakage-free logical unitary $R=4N$; at $n=3$ states need $R\approx3N$ and unitaries are not reached up to $R=6N$.}
\label{fig:reach}
\end{figure}

\begin{table}[t]
\caption{State-preparation fidelity (mean over 6 Haar-random and 6 random $\chi=2$ MPS logical targets, best of 3 L-BFGS restarts) for bubble-sort ($R$ rounds) and staircase ($L$ layers) networks. $p$ = number of pulses. \texttt{st}: $n=3$, $N=6$, initial state $\ket{S T_0 S}$; \texttt{eo}: $n=2$, $N=6$, initial state $\ket{0_L0_L}$.}
\label{tab:reach}
\begin{ruledtabular}
\begin{tabular}{llrccc}
enc & network & $p$ & $F$ (Haar) & $F$ (MPS) & $F_{\max}$ (Haar)\\
\hline
st & bubble $R{=}6$ & 15 & 0.366 & 0.410 & 0.381\\
st & bubble $R{=}12$ & 30 & 0.381 & 0.426 & 0.381\\
st & stair $L{=}3$ & 15 & 0.359 & 0.406 & 0.381\\
eo & bubble $R{=}6$ & 15 & 0.951 & 0.976 & 1\\
eo & bubble $R{=}7$ & 18 & 0.9986 & 0.9999 & 1\\
eo & bubble $R{=}8$ & 20 & $1-10^{-8}$ & $1-10^{-8}$ & 1\\
eo & stair $L{=}3$ & 15 & 0.931 & 0.963 & 1\\
eo & stair $L{=}4$ & 20 & $1-10^{-8}$ & $1-10^{-8}$ & 1\\
\end{tabular}
\end{ruledtabular}
\end{table}

\subsection{Selection rule for PSB readout}
\label{sec:rule}
\begin{figure}[t]
\includegraphics[width=\columnwidth]{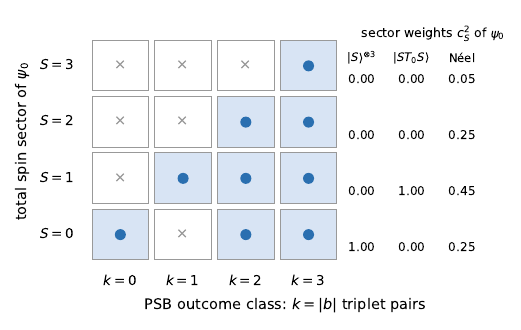}
\caption{The selection rule for the two-spin encoding with $n=3$ blocks ($N=6$). A dot marks the total-spin sectors $S\in\Lambda(k)$ of the initial state that can contribute to a PSB outcome with $k$ triplet pairs; a cross marks a forbidden combination (one spin-1 cannot couple to $S=0$, so $k=1$ is fed by $S=1$ only, and $k$ spin-1's cannot exceed $S=k$). Exchange dynamics cannot move weight between rows, so the probability of a column is bounded by the weights $c_S^2$ of its dotted rows, Eq.~(\ref{eq:rule}). Right: $c_S^2$ for three initial states. $\ket S^{\otimes3}$ lives in $S=0$ and can never produce $k=1$; $\ket{ST_0S}$ (one $T_0$ between two singlets) lives in $S=1$ and can never produce the all-singlet outcome $k=0$; the N\'eel state has weight in every row and forbids no outcome class (its floor, Eq.~(\ref{eq:tvfloor}), is zero for both six-qubit teachers of Sec.~\ref{sec:distil}; the row weights still cap each column).}
\label{fig:sectors}
\end{figure}

For distillation the relevant ceiling is on distributions, not states. Consider the two-spin encoding, $N=2n$, $S_z=0$, and a PSB outcome $b$ with $k=|b|$ triplet pairs. The image of $\Pi_b$ consists of states that are singlets on $n-k$ pairs and spin-1 on the remaining $k$; coupling $k$ spin-1's gives total spin $S\in\Lambda(k)$ with
\begin{equation}
\Lambda(k)=\{0,1,\dots,k\}\ (k\ne1),\qquad \Lambda(1)=\{1\},
\end{equation}
since one spin-1 cannot couple to $S=0$ (Fig.~\ref{fig:sectors}). Therefore $\Pi_bP_S=0$ for $S\notin\Lambda(|b|)$.

 For any set $K$ of Hamming classes, writing $\Lambda(K)=\bigcup_{k\in K}\Lambda(k)$ and $\psi=\mathcal B\psi_0$, this gives $\Pi_b\psi=\Pi_b\sum_{S\in\Lambda(K)}P_S\psi$ for $|b|\in K$; the $\Pi_b$ are mutually orthogonal projectors, and $\mathcal B$ commutes with every $P_S$, so
\begin{equation}
P(|b|\in K)=\sum_{|b|\in K}\Bigl\|\Pi_b\!\!\sum_{S\in\Lambda(K)}\!\!P_S\psi\Bigr\|^2\le\sum_{S\in\Lambda(K)}c_S^2,
\label{eq:rule}
\end{equation}
with $c_S^2=\|P_S\psi_0\|^2$ the sector weights of the \emph{initial} state, which exchange dynamics cannot change. Consequently, for any teacher distribution $p_t$,
\begin{equation}
\mathrm{TV}(p_t,p)\ \ge\ \max_K\Bigl[\,p_t(|b|\in K)-\sum_{S\in\Lambda(K)}c_S^2\Bigr]_+
\label{eq:tvfloor}
\end{equation}
with $[x]_+=\max(x,0)$, for every exchange-only circuit of any depth (Eq.~(\ref{eq:tvfloor}) is $\mathrm{TV}\ge p_t(A)-p(A)$ applied to $A=\{b:|b|\in K\}$ with Eq.~(\ref{eq:rule})). Special cases of the rule have been used in experiments: from a product of local singlets the outcomes $ST$ and $TS$ of a four-dot chain are forbidden \cite{vanDiepen2021}, which is $\Lambda(1)=\{1\}$ applied to $S=0$, and our earlier work \cite{Minato2026} noted that the N\'eel state has support on every total-spin block; Eq.~(\ref{eq:tvfloor}) makes the constraint quantitative for arbitrary $n$ and targets.

For the three-spin encoding the same argument gives $S\in\Lambda(k)\otimes\{\text{$n$ spin-$\tfrac12$'s}\}$, which always contains the logical sector $S=n/2$; starting from a code state, Eq.~(\ref{eq:rule}) is vacuous and there is no structural floor from the symmetry (Sec.~\ref{sec:suff}).

\subsection{The rule is also sufficient}
\label{sec:suff}
Equation~(\ref{eq:rule}) is not only necessary. Because $\Pi_b$ is SU(2)-invariant it commutes with every $P_S$, so for $\psi=\sum_Sc_S\phi_S$ ($\phi_S$ a unit vector in sector $S$ at $S_z=0$)
\begin{equation}
p(b)=\sum_S c_S^2\,x_{S,b},\qquad x_{S,b}=\|\Pi_b\phi_S\|^2,
\label{eq:flow}
\end{equation}
with no cross terms between sectors: the relative phases of the $c_S$ drop out. For fixed $S$ the subspaces $V_{S,b}=\Pi_bP_S\mathcal H$ are mutually orthogonal and sum to $P_S\mathcal H$, and $V_{S,b}\ne0$ exactly when $S\in\Lambda(|b|)$ (for every $S\in\Lambda(k)$ there is an $S_z=0$ state of $k$ spin-1's with total spin $S$). Hence, as $\phi_S$ ranges over the unit sphere of $P_S\mathcal H$, the row $x_{S,\cdot}$ ranges over \emph{all} probability distributions supported on $\{b:\ S\in\Lambda(|b|)\}$, independently for each $S$. Exchange gates on a chain generate, in the closure, every unitary of the form $\bigoplus_SU_S$ with $U_S\in SU(m_S)$ \emph{simultaneously and independently} for all $S$ (semi-universality; the only restriction is on the relative phases between sectors) \cite{Marvian2024,Hulse2024,Kempe2001}, and $SU(m_S)$ acts transitively on the unit sphere of $P_S\mathcal H$ when $m_S\ge2$ (when $m_S=1$---at $S_z=0$ only the top sector $S=n$---the row $x_{S,\cdot}$ is fixed, but that sector is allowed only the single outcome $b=1\cdots1$, so nothing is lost). Hence every collection of rays $(\phi_S)_S$ is reachable in the closure, and for the readout problem the sector norms $c_S$ are the only invariants. The construction is explicit: given rows $x_{S,\cdot}$, pick unit vectors $v_{S,b}\in V_{S,b}$ and set $\phi_S=\sum_b\sqrt{x_{S,b}}\,v_{S,b}$.

The reachable readout distributions are therefore exactly the marginals $p(b)=\sum_Sc_S^2x_{S,b}$ of a transportation problem: supplies $c_S^2$ at the sectors, demands $p(b)$ at the outcomes, and an edge $S$--$b$ whenever $S\in\Lambda(|b|)$. By the max-flow/min-cut theorem a target $p_t$ is reachable if and only if $p_t(A)\le\sum_{S\in\Lambda(A)}c_S^2$ for every set $A$ of outcomes, where $\Lambda(A)$ is the union of $\Lambda(|b|)$ over $b\in A$. If $K$ is the set of Hamming classes present in $A$ and $B_K=\{b:|b|\in K\}$, then $\Lambda(A)=\Lambda(B_K)$ and $p_t(A)\le p_t(B_K)$, so the tightest constraints are the unions of classes, and the condition is Eq.~(\ref{eq:rule}). The same cut gives the value of the maximum flow, $F^*=1-\max_K[\,p_t(|b|\in K)-\sum_{S\in\Lambda(K)}c_S^2]_+$, and $\mathrm{TV}(p_t,p)=1-\sum_b\min[p_t(b),p(b)]\ge1-F^*$ for every reachable $p$, because capping each column of the transportation plan of $p$ at $p_t(b)$ gives a feasible flow of value $\sum_b\min[p_t(b),p(b)]$. Equality is attained: in a maximum flow every outcome adjacent to a sector with unused supply is saturated (otherwise the flow could be augmented), so the unused supply can be placed on such outcomes without lowering the overlap. In short: \emph{the closure of the set of PSB distributions reachable by exchange-only circuits from $\psi_0$ is the polytope cut out by Eq.~(\ref{eq:rule}), and the minimum of $\mathrm{TV}(p_t,p)$ over it equals the right-hand side of Eq.~(\ref{eq:tvfloor}).} The statement concerns the closure at unbounded depth, i.e.\ an infimum over finite circuits; what the bubble-sort network reaches at a given depth $R$ is the subject of Sec.~\ref{sec:distil}.

For the three-spin encoding ($N=3n$, $S_z=n/2$) from a code state all the weight sits in $S=n/2$, and $V_{n/2,b}\ne0$ for every $b$: a block whose first pair is read as a singlet contains $\ket S\ket\uparrow$, one whose first pair is read as a triplet contains $\sqrt{2/3}\ket{T_+}\ket\downarrow-\sqrt{1/3}\ket{T_0}\ket\uparrow$, both of block spin $\tfrac12$, $m=\tfrac12$, and the product of $n$ such highest-weight doublets has $S=n/2$. The polytope is then the full simplex and there is no floor from the symmetry.

\subsection{Numerical illustration}
\label{sec:tight}
Before the argument above was found we had tested Eq.~(\ref{eq:tvfloor}) numerically by discarding the circuit and minimising $\mathrm{KL}(p_t\|p)$ directly over $\{\sum_Sc_S\phi_S\}$ (924-dimensional $S_z=0$ sector of $N=12$, Adam, 1500 steps, 2--3 seeds agreeing to $10^{-4}$), for the two GPT-2 teachers of Sec.~\ref{sec:teach}. Table~\ref{tab:rule} lists the sector weights of several product initial states, the floor of Eq.~(\ref{eq:tvfloor}), and the TV at the numerical optimum; in all ten cases (seven with a non-zero floor) they agree to four digits, consistent with the theorem (the search minimised KL, not TV, so agreement was not guaranteed a priori). The test checks the transportation polytope; that exchange circuits reach it is the semi-universality theorem cited above, not something the numerics confirm. When the sector weights themselves are made free parameters the optimum is $\mathrm{KL}=\mathrm{TV}=0$ (with non-unique weights), confirming that the limitation of the two-spin encoding is entirely in the initial state.

\begin{table}[t]
\caption{Selection-rule floors for the two-spin encoding, $n=6$ ($N=12$). $c_S^2$: weights of the initial product state in total-spin sectors $S=0,\dots,6$ (multiplicities $132,297,275,154,54,11,1$ at $S_z=0$). Floor: Eq.~(\ref{eq:tvfloor}). $\ket{ST_0}^{\otimes3}$ is the alternating state ``alt''. Opt: TV at the minimum of KL over the full symmetry-allowed set. Teacher \texttt{cat}: $p_t(|b|=0)=0.0117$, $p_t(|b|=1)=0.0528$, $p_t(|b|\le3)=0.781$; teacher \texttt{begin}: $p_t(|b|=0)=0$, $p_t(|b|=1)=0.248$.}
\label{tab:rule}
\footnotesize
\begin{ruledtabular}
\begin{tabular}{l@{\hspace{3pt}}l@{\hspace{3pt}}c@{\hspace{5pt}}c@{\hspace{5pt}}c@{\hspace{5pt}}c}
initial state & $c_S^2$ ($S=0..6$) & \multicolumn{2}{c}{\texttt{cat}} & \multicolumn{2}{c}{\texttt{begin}}\\
 & & floor & opt & floor & opt\\
\hline
$\ket{S}^{\otimes6}$ & 1,0,0,0,0,0,0 & .0528 & .0528 & .248 & .248\\
$\ket{ST_0}^{\otimes3}$ & 0,.60,0,.40,0,0,0 & .0117 & .0117 & 0 & 0\\
$\ket{T_0}^{\otimes6}$ & .14,0,.48,0,.31,0,.07 & .162 & .162 & .248 & .248\\
$\ket{ST_0S}^{\otimes 2}$ & .33,0,.67,0,0,0,0 & .0528 & .0528 & .248 & --\\
$\ket{ST_0T_0}^{\otimes 2}$ & .20,0,.57,0,.23,0,0 & .0528 & .0528 & .248 & --\\
$\ket{\uparrow\downarrow}^{\otimes6}$ (N\'eel) & .14,.32,.30,.17,.06,.01,.001 & 0 & 0 & 0 & 0\\
free $c_S$ & optimised & 0 & 0 & -- & --\\
\end{tabular}
\end{ruledtabular}
\end{table}

The physics is simple. A product of singlets and $T_0$'s with $m$ triplet pairs has total spin of parity $(-1)^m$ only (Table~\ref{tab:rule}): it is a product of $m_z=0$ states of spins $j=0$ and $1$, and the Clebsch--Gordan coefficient $\braket{j_1\,0;j_2\,0|J\,0}$ vanishes when $j_1+j_2+J$ is odd, so at least one Hamming class of outcomes is inaccessible: with $\ket{ST_0ST_0\cdots}$ the all-singlet outcome $000000$ can never be observed ($S=0$ is absent), and with $\ket{S}^{\otimes6}$ no outcome with exactly one triplet can ($S=1$ is absent). The N\'eel state $\ket{\uparrow\downarrow}^{\otimes 6}=\bigotimes_j(\ket S+\ket{T_0})/\sqrt2$, among the simplest states to prepare on a dot array, contains every sector, so that no outcome class is forbidden and the floor for the two primary teachers is below $10^{-6}$ (Table~\ref{tab:rule}; for other teachers the capacity floor can be large, Sec.~\ref{sec:other}); that a computational-basis product state has weight in every total-spin sector compatible with its magnetisation is Lemma~6 of Ref.~\cite{Burkat2026}. (The $k=0$ and $k=2$ classes inside $S=0$ are the singlet-only qubit of Ref.~\cite{Sala2017}.) For the three-spin encoding the code state $\ket{0_L}^{\otimes n}$ is already in the unique logical sector.

\section{Distilling GPT-2}
\label{sec:distil}

\subsection{Teachers and MPS compression}
\label{sec:teach}
We take GPT-2 small \cite{Radford2019}, fix a prompt, restrict the vocabulary to the top-$V$ next tokens, and generate the chain distribution $p_t(t_1,\dots,t_L)$ by renormalising the restricted softmax at each step. With $V=8,L=2$ or $V=4,L=3$ the outcome space has 64 elements, encoded as six bits. Two teachers are used: \texttt{cat} (``The cat sat on'', $V=8$, $L=2$, entropy 3.78 bit) and \texttt{begin} (``In the beginning'', $V=4$, $L=3$, 3.45 bit).

Each teacher is first fit by an MPS Born machine $p(b)=|\braket{b|\Psi_{\rm MPS}}|^2$ of bond dimension $\chi$ (Adam, 3000 steps, best of 3 seeds): $\mathrm{KL}(p_t\|p)$ is $0.351/0.136/0.042/0.005$ nat for \texttt{cat} and $0.729/0.168/0.010/0.001$ for \texttt{begin} at $\chi=1/2/3/4$, and zero at $\chi=8$. The MPS is the control for the tensor-network stage of the pipeline, not a strong classical baseline (a 64-outcome distribution is represented exactly by 63 real parameters), and it is given a larger optimisation budget than the exchange network below (3000 steps and 3 seeds against 300--400 steps and a single run per initialisation), so comparisons with it are conservative for the network. No claim of quantum advantage is made in this paper. These values serve as the scale on which the exchange network is judged, in the sense of Rudolph \emph{et al.} \cite{Rudolph2023}, where the bond dimension measures the classical resource a circuit must emulate: we take the $\chi=4$ level as the operational criterion for a network that has learned the teacher, and the $\chi=1$ level for one that has learned only its marginals. MPS Born machines are themselves sequence models \cite{Miller2021}, tensor networks are used to compress the weights of large language models \cite{Tomut2024} and of classical vision networks \cite{Xiafukaiti2024}, and matrix product states have been used to analyse recurrent language models \cite{Tangpanitanon2022}; quantum Born machines have been trained on corpus bigram statistics \cite{Widdows2024} and language models have been distilled into variational circuits for classification \cite{Li2025QD}, but we are not aware of a distillation of a language model's generative distribution into an exchange-only circuit, with MPS Born machines as the classical control. Hybrid quantum language models with a classical readout head have been trained on hardware \cite{Balauca2025}; here the register's Born distribution \emph{is} the token distribution, with no classical head. MPS Born machines with trainable token embeddings have been compared with GPT-2 as sequence models \cite{Hou2023}, and quantum sequence models are provably more expressive than Bayesian networks of the same size \cite{Gao2022}. We train on the exact KL divergence rather than on a moment-matching loss, since MMD-trained generative models transferred from classical simulation generalise poorly \cite{Raj2026}.

\subsection{Training the exchange network}
The network parameters $\theta$ are trained by gradient descent on the smoothed divergence $\mathrm{KL}_\epsilon(p_t\|p_\theta)=\sum_bp_t(b)\,[\ln(p_t(b)+\epsilon)-\ln(p_\theta(b)+\epsilon)]$ with $\epsilon=10^{-12}$ and no renormalisation (natural logarithm throughout; entropies are quoted in bits); it agrees with $\mathrm{KL}(p_t\|p_\theta)$ to $O(\epsilon/p)$ wherever $p_\theta\gg\epsilon$ (and can be slightly negative, by less than $10^{-7}$, because the stored teacher tables are normalised only to $10^{-7}$; floors and divergences below $10^{-6}$ are reported as zero throughout), and caps the contribution of a structurally zero outcome at $p_t(b)\ln[p_t(b)/\epsilon]$, which is why finite values are reported for initial states with a floor (for those, TV is the quantity the theory constrains and is always given alongside). Here $p_\theta(b)=\|\Pi_b\mathcal B_R(\theta)\psi_0\|^2$ (state-vector simulation, $2^{12}$ or $2^{18}$ amplitudes, Adam, learning rate $0.05$, 300--400 steps; PyTorch, CPU). Four initialisations are compared: random pulse areas in $[0,2\pi)$, $\theta\equiv0$ and $\theta\equiv\pi$ (the SWAP network), both with a $0.1$-rad Gaussian jitter, and an MPS pretraining in which the network is first fitted by L-BFGS to the embedded $\chi=4$ MPS state (Rudolph \emph{et al.} \cite{Rudolph2023}) and then fine-tuned on the KL. Leakage is neither penalised nor constrained.

\begin{figure}[t]
\includegraphics[width=\columnwidth]{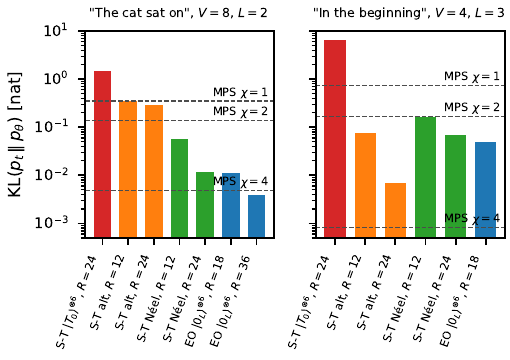}
\caption{Final KL divergence (nat) of the trained exchange network to the two GPT-2 teachers, best over pulse initialisations and step counts, for the two-spin encoding with three initial states ($\ket{T_0}^{\otimes6}$, alternating singlet/triplet, N\'eel) and the three-spin encoding from $\ket{0_L}^{\otimes6}$. Dashed lines: MPS Born machines with bond dimension $\chi=1,2,4$. The $\ket{T_0}^{\otimes6}$ bar is dominated by structural zeros of the selection rule (Sec.~\ref{sec:rule}); the N\'eel state removes the floor for both teachers but only wins for the teacher whose floor was non-zero.}
\label{fig:distil}
\end{figure}

\subsection{Results}
Table~\ref{tab:distil} and Fig.~\ref{fig:distil} collect the main runs. Three findings:

\begin{table*}[t]
\caption{Distillation of GPT-2 teacher distributions (64 outcomes, six logical qubits) into bubble-sort ($R$) and staircase ($L$) exchange networks. $p$ = pulses. KL in nat, TV = total variation, leak = weight of the final state outside the code space. MPS reference for \texttt{cat}: $\chi=1$: 0.351, $\chi=2$: 0.136, $\chi=4$: 0.005; for \texttt{begin}: 0.729 / 0.168 / 0.001. Initial states: alt $=\ket{ST_0ST_0\cdots}$, N\'eel $=\ket{\uparrow\downarrow}^{\otimes n}$, $\ket{0_L}^{\otimes n}$ for \texttt{eo}; init: pulse-area initialisation (random, $0$, $\pi$, or MPS pretraining with the fidelity reached to the embedded MPS state). KL means $\mathrm{KL}_\epsilon$.}
\label{tab:distil}
\begin{ruledtabular}
\begin{tabular}{lllrlccc}
teacher & enc ($N$) & initial state & network & $p$ & init & KL & TV / leak\\
\hline
cat & st (12) & alt & bubble $R{=}12$ & 66 & random & 0.377 & 0.20 / 0.89\\
cat & st (12) & alt & bubble $R{=}12$ & 66 & $\pi$ & 0.476 & 0.28 / 0.80\\
cat & st (12) & alt & bubble $R{=}24$ & 132 & random & 0.333 & 0.17 / 0.90\\
cat & st (12) & alt & bubble $R{=}24$ & 132 & random, 2000 steps & 0.284 & 0.10 / 0.90\\
cat & st (12) & alt & stair $L{=}6$ & 66 & random & 0.388 & 0.21 / 0.90\\
cat & st (12) & alt & bubble $R{=}12$ & 66 & MPS$_4$ pretrain ($F{=}0.24$) & 0.374 & 0.20 / 0.87\\
cat & st (12) & $\ket{T_0}^{\otimes6}$ & bubble $R{=}24$ & 132 & random & 1.47 & 0.36 / 0.61\\
cat & st (12) & N\'eel & bubble $R{=}12$ & 66 & random & 0.117 & 0.19 / 0.84\\
cat & st (12) & N\'eel & bubble $R{=}12$ & 66 & $\pi$ & 0.058 & 0.11 / 0.82\\
cat & st (12) & N\'eel & stair $L{=}6$ & 66 & random & 0.082 & 0.15 / 0.82\\
cat & st (12) & N\'eel & bubble $R{=}6$ & 33 & $\pi$ & 0.250 & 0.31 / 0.80\\
cat & st (12) & N\'eel & bubble $R{=}18$ & 99 & $\pi$ & 0.062 & 0.12 / 0.83\\
cat & st (12) & N\'eel & bubble $R{=}24$ & 132 & random & 0.034 & 0.08 / 0.84\\
cat & st (12) & N\'eel & bubble $R{=}24$ & 132 & $\pi$ & 0.014 & 0.04 / 0.83\\
cat & st (12) & N\'eel & bubble $R{=}36$ & 198 & $\pi$ & \textbf{0.0029} & 0.02 / 0.79\\
cat & eo (18) & $\ket{0_L}^{\otimes6}$ & bubble $R{=}18$ & 153 & random & 0.0136 & 0.038 / 0.95\\
cat & eo (18) & $\ket{0_L}^{\otimes6}$ & bubble $R{=}18$ & 153 & $\pi$ & 0.0112 & 0.034 / 0.92\\
cat & eo (18) & $\ket{0_L}^{\otimes6}$ & bubble $R{=}18$ & 153 & 0 & 0.202 & 0.22 / 0.90\\
cat & eo (18) & $\ket{0_L}^{\otimes6}$ & bubble $R{=}18$ & 153 & MPS$_4$ pretrain ($F{=}0.76$) & 0.024 & 0.050 / 0.61\\
cat & eo (18) & $\ket{0_L}^{\otimes6}$ & stair $L{=}9$ & 153 & random & 0.018 & 0.043 / 0.96\\
cat & eo (18) & $\ket{0_L}^{\otimes6}$ & bubble $R{=}36$ & 306 & random & \textbf{0.0038} & 0.019 / 0.95\\
cat & eo (18) & $\ket{0_L}^{\otimes6}$ & bubble $R{=}36$ & 306 & MPS$_4$ pretrain ($F{=}0.88$) & 0.0051 & 0.025 / 0.55\\
\hline
begin & st (12) & alt & bubble $R{=}12$ & 66 & random & 0.114 & 0.14 / 0.68\\
begin & st (12) & alt & bubble $R{=}24$ & 132 & random & 0.032 & 0.06 / 0.69\\
begin & st (12) & alt & bubble $R{=}24$ & 132 & random, 2000 steps & 0.0068 & 0.020 / 0.66\\
begin & st (12) & alt & bubble $R{=}36$ & 198 & random & 0.0136 & 0.028 / 0.66\\
begin & st (12) & $\ket{T_0}^{\otimes6}$ & bubble $R{=}24$ & 132 & random & 6.45 & 0.30 / 0.87\\
begin & st (12) & N\'eel & bubble $R{=}12$ & 66 & random & 0.226 & 0.21 / 0.71\\
begin & st (12) & N\'eel & bubble $R{=}12$ & 66 & $\pi$ & 0.161 & 0.18 / 0.67\\
begin & st (12) & N\'eel & bubble $R{=}24$ & 132 & random & 0.074 & 0.11 / 0.66\\
begin & st (12) & N\'eel & bubble $R{=}36$ & 198 & $\pi$ & 0.0275 & 0.06 / 0.66\\
begin & eo (18) & $\ket{0_L}^{\otimes6}$ & bubble $R{=}18$ & 153 & random & 0.049 & 0.063 / 0.99\\
begin & eo (18) & $\ket{0_L}^{\otimes6}$ & bubble $R{=}18$ & 153 & MPS$_4$ pretrain ($F{=}0.90$) & 0.059 & 0.071 / 0.88\\
\end{tabular}
\end{ruledtabular}
\end{table*}

\emph{(a) The three-spin encoding comes within a factor of three of the $\chi=4$ level; the two-spin encoding from products of singlets and $T_0$'s stalls between the $\chi=1$ and $\chi=2$ levels.} At the same number of rounds $R=N$, \texttt{eo} reaches $\mathrm{KL}=0.011$--$0.014$ for \texttt{cat} (within a factor of three of the $\chi=4$ MPS, TV $0.035$) and $0.049$ for \texttt{begin} (between $\chi=2$ and $\chi=4$; this teacher has lower entropy, $3.45$ against $3.78$ bit, and its $\chi=4$ MPS happens to be better, $0.001$ against $0.005$), while \texttt{st} from $\ket{ST_0ST_0ST_0}$ stops at $0.33$--$0.48$ for \texttt{cat} (between $\chi=1$ and $\chi=2$) and at $0.11$ for \texttt{begin}. Doubling the depth, changing to a staircase, extending training to 2000 steps ($0.377\to0.346$, $0.333\to0.284$), or pretraining on the MPS does not move \texttt{st} appreciably for \texttt{cat}. From $\ket{T_0}^{\otimes6}$ the result is worse still ($1.5$; $6.4$ for \texttt{begin}).

\begin{figure}[t]
\includegraphics[width=\columnwidth]{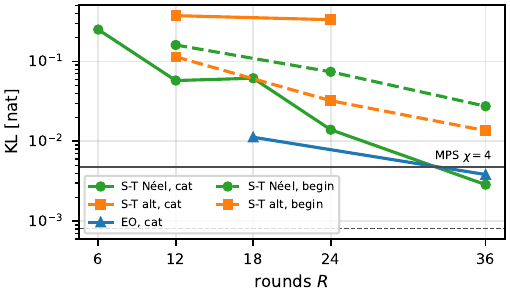}
\caption{KL divergence versus number of rounds for the two-spin encoding with the alternating and the N\'eel initial state (400 Adam steps for the two-spin runs, 300 for the three-spin, best over pulse initialisations, seed 0; solid: \texttt{cat}, dashed: \texttt{begin}) and for the three-spin encoding where available. Horizontal lines: $\chi=4$ MPS. With a structural floor (alt, \texttt{cat}) depth does not help; without one, the KL falls roughly geometrically with $R$.}
\label{fig:depth}
\end{figure}

\emph{(b) The selection rule accounts for the ordering, and the N\'eel state repairs it.} Table~\ref{tab:rule} predicts floors of $0.012$ (alt) and $0.16$ (all $T_0$) in TV for \texttt{cat} and $0$ / $0.25$ for \texttt{begin}, matching the ranking in Table~\ref{tab:distil}; the KL values are inflated by the structurally zero outcomes (for alt, the single term $p_t(000000)\ln[p_t(000000)/\epsilon]\approx0.27$ nat). Simply initialising the same chain in $\ket{\uparrow\downarrow}^{\otimes6}$ lowers the KL for \texttt{cat} by a factor 6--25 at equal depth ($0.377\to0.058$ at $R=12$, $0.333\to0.014$ at $R=24$, from the $\pi$ initialisation, which is consistently the better start for the N\'eel state), and, unlike the alternating state, keeps improving with depth (Fig.~\ref{fig:depth}): $0.25,\,0.058,\,0.062,\,0.014,\,0.0029$ at $R=6,12,18,24,36$ ($R=12$ and $18$ are level within the 400-step budget). These are single runs (seed 0); over five seeds the $R=24$ network gives a median of $0.014$ (range $0.012$--$0.017$) and the $R=36$ network $0.0038$ ($0.0024$--$0.0080$), so the depth ordering is robust while the last digit is not. At $R=36$ (198 pulses on 12 dots) the two-spin chain is below the $\chi=4$ MPS ($0.005$) and below the three-spin chain at $R=18$ ($0.011$, 153 pulses on 18 dots); the three-spin chain at $R=36$ (306 pulses) reaches $0.0038$, so at equal depth the two encodings end within a factor $1.3$ of each other and both below the MPS that was used as the classical control. For \texttt{begin}, whose all-singlet outcome has teacher probability $4\times10^{-7}$, the alternating initial state has a floor of the same size, i.e.\ none in practice, and there it does as well as or better than N\'eel ($0.032$ versus $0.074$ at $R=24$; N\'eel over five seeds: median $0.082$, range $0.069$--$0.106$): the N\'eel state is the safe choice because it has no forbidden outcome class, so that its floor is set only by the sector capacities (negligible, $10^{-9}$ for \texttt{cat}, for the teachers studied), not because it is uniformly the best. The remaining gap to the floor is partly optimisation and partly expressivity, and the split depends on the teacher: for \texttt{begin}-alt (floor $0$) training the $R=24$ network for 2000 instead of 400 steps takes the KL from $0.032$ to $0.0068$ (TV $0.020$), about eight times the $\chi=4$ MPS, whereas for \texttt{cat}-alt the same extension moves the KL only from $0.333$ to $0.284$ because the structural zero fixes the KL from below. Where no floor exists the two-spin network keeps improving with training; where one exists training cannot go below it. The two-spin encoding therefore does work as a distillation target, provided its initial state is chosen by the selection rule rather than by the code.

\emph{(c) The leakage space is a shortcut.} In every run the final state has $35$--$99\%$ of its weight outside the code space ($55$--$99\%$ for the two primary teachers), with $0.92$--$0.95$ in the best \texttt{eo} runs and up to $0.99$ in worse ones, so leakage is not by itself predictive of the KL. Nothing in the objective pulls the state back; the readout is defined on the full space, and the path through---and destination in---the leakage space accompanies every good solution we found; whether it is necessary for the expressivity is not tested here (a leakage-penalised control would be needed). This is the intended use of ``leakage'' here: it is a resource for a Born machine, and an error only for a gate-model computation that must return to the code space, where it is removed by dedicated pulse sequences \cite{Fong2011}. That PSB sees the leaked states is what makes this consistent: in blind randomised benchmarking of a triple-dot qubit a triplet outcome may equally be a logical $\ket{1_L}$ or a leaked $S=\tfrac32$ state \cite{Andrews2019,Blumoff2022}, and here the two are deliberately not distinguished. (The projectors $\Pi_{\rm s}$ and $\mathbb 1-\Pi_{\rm s}$ are themselves SU(2)-invariant; the readout is the only non-unitary step of the scheme. Measurement-based schemes that reach universality from exchange and singlet--triplet measurements \cite{Brooks2021} need further resources beyond a terminal readout.)

\emph{Initialisation.} For \texttt{eo}, $\theta\equiv\pi$ (the pure SWAP network) is as good as or slightly better than random ($0.011$ versus $0.014$), and gives a much better starting KL for \texttt{st}-N\'eel ($1.48$ versus $3.0$). Both $\theta\equiv0$ and $\theta\equiv\pi$ are exact stationary points of the KL for every real product initial state: the state stays real up to a global phase, each pulse generator is real symmetric, so $\partial p(b)/\partial\theta_k=2\,\mathrm{Re}\braket{\psi|\Pi_b|\partial_k\psi}$ is the real part of a purely imaginary number and vanishes (checked to machine precision for all three initial states); the $0.1$-rad jitter we add is what moves the optimiser off them. From there the two starts are not equivalent. At $\theta\equiv0$ the readout of $\ket{0_L}^{\otimes6}$ is the deterministic outcome $000000$, to which the teacher assigns $1.2\%$ (KL $5.8$ after jitter), and the optimiser falls into a poor optimum ($0.20$, twenty times worse than $\pi$); at $\theta\equiv\pi$ the reversal maps every block onto a block with its internal order reversed, so the pair that is read out is a different pair of the original block (starting KL $2.85$, the same as a random network), and the optimiser proceeds from there. For \texttt{st} with singlet/triplet initialisation the three initialisations differ by $<0.1$ (except from $\ket{T_0}^{\otimes6}$, where they span $0.26$ around $1.8$ for \texttt{cat} and $0.35$ around $6.7$ for \texttt{begin}). We do not observe barren-plateau behaviour at these sizes and do not claim anything about it; the $S_z$ and $\mathbf S^2$ conservation of the network reduces the effective dimension \cite{Larocca2022,Schatzki2024,Nguyen2024} (the dynamical Lie algebra of the nearest-neighbour Heisenberg brickwork is classified in Ref.~\cite{Wiersema2024}), which is expected to help at larger $N$ but is not tested here. MPS pretraining \cite{Rudolph2023,Dborin2022} does not help either encoding. For \texttt{eo} the network at $R=N$ can only reach fidelity $0.76$ to the embedded $\chi=4$ state (six-qubit state preparation was not attempted; two and three qubits needed $R\approx N+2$ to $3N$, Sec.~\ref{sec:ceiling}), which gives a good starting KL ($0.42$) but a worse final one ($0.024$ versus $0.014$ from random) and a markedly less leaked final state ($0.61$ versus $0.92$--$0.95$); at $R=2N$ the pretraining fidelity rises to $0.88$ and the ordering is the same ($0.0051$ versus $0.0038$, leakage $0.55$ versus $0.95$); for \texttt{begin} at $R=N$, $0.059$ versus $0.049$: in these runs, pretraining towards a code-space state starts the optimiser in a region from which it does not reach the leaked solutions found from random starts (we did not test whether the leakage itself is what helps). For \texttt{st} the pretraining fidelity is capped at $0.24$--$0.30$ by the symmetry ceiling and the fine-tuned KL equals the random-start value.

\subsection{Pulse budget}
The \texttt{eo} network that comes within a factor of three of the $\chi=4$ level uses 153 exchange pulses in 18 layers on 18 dots (the one that passes it, at $R=2N$, uses 306). A gate-model preparation of a $\chi=4$ MPS on six EO qubits needs, in the sequential construction \cite{Schon2005,Ran2020}, a staircase of three-qubit isometries (bond dimension $4$ is two qubits of bond), each of which decomposes into several CNOTs at 18--22 exchange pulses apiece in the Fong--Wandzura construction \cite{Fong2011,Zeuch2016} plus single-qubit layers; a two-qubit staircase prepares only $\chi=2$ exactly. Either way the count would be of order a few hundred leakage-free pulses at comparable or larger depth (we have not carried out this compilation). The comparison changes the target, the method and the leakage constraint at once, so it bounds the gate-model cost rather than isolating the cause of the difference; a controlled comparison would train the same network with and without a leakage penalty at matched accuracy, which we have not done.

\emph{Schedule.} At equal pulse count (153) the staircase schedule with $L=9$ layers is slightly worse than the bubble-sort schedule with $R=18$ rounds for \texttt{eo} ($0.018$ versus $0.011$--$0.014$) and equal for \texttt{st} at 66 pulses ($0.388$ versus $0.377$; N\'eel: $0.082$ versus $0.058$--$0.117$); the difference between the two schedules is small compared with the difference between initial states, but the bubble-sort schedule has depth $R$ against the staircase's $L(N-1)$, i.e.\ 18 against 153 layers here.

\subsection{Is the chain the bottleneck?}
\label{sec:connect}
Property (v) of Sec.~\ref{sec:network} says that no layout brings every pair of spins together in fewer than $N-1$ rounds; it does not say that a chain optimises as well as a richer layout at the same depth. We therefore repeated the distillation---same pulses, readout, loss, training and initial states---on four schedules of disjoint pulses at equal depth $R$: the chain (odd--even transposition); a two-dimensional array of $3\times4$, $4\times4$ or $3\times6$ dots with the logical blocks along the rows and the rounds cycling through row-even, column-even, row-odd and column-odd pairs; a non-local round robin (the circle method: $N-1$ perfect matchings of $K_N$ that cover every pair exactly once, so every pair of dots interacts directly); and a fresh random perfect matching of all $N$ dots in every round. The matchings have $N/2$ pulses per round, the chain alternates between $N/2$ and $N/2-1$, and the array has $20$--$23\%$ fewer; the comparison is at fixed depth, the hardware-relevant quantity, and a pulse-matched comparison would favour the array somewhat. Table~\ref{tab:connect} gives, for each seed, the better of the $\pi$ and random initialisations, with the median and range over five seeds (\texttt{st}) or three (\texttt{eo}) at the depths $R=N$ and $2N$ that carry the conclusions.

\begin{table*}[t]
\caption{KL divergence (nat) at equal depth $R$ for four schedules of disjoint exchange pulses: the linear chain (odd--even transposition), a two-dimensional array, a non-local round robin of all pairs, and random perfect matchings. For each seed the better of the $\pi$ and random initialisations is taken; entries with five (\texttt{st}) or three (\texttt{eo}) seeds give the median and, in brackets, the range; single-seed entries are marked $^{1}$. 400 Adam steps (300 for \texttt{eo}). Teachers \texttt{cat} ($n=6$: 64 outcomes; $n=8$: 256 outcomes); N\'eel initial state for \texttt{st}, $\ket{0_L}^{\otimes n}$ for \texttt{eo}. Reference $\chi=4$ MPS: $0.005$ ($n=6$), $0.036$ ($n=8$).}
\label{tab:connect}
\begin{ruledtabular}
\begin{tabular}{lcllll}
enc./$n$ & $R$ & chain & array & r.\,robin & random\\
\hline
st/6 & $N/2$ & 0.250$^{1}$ & 0.264$^{1}$ & 0.228$^{1}$ & 0.263$^{1}$\\
st/6 & $N$ & 0.058 [.047--.071] & 0.094 [.084--.128] & \textbf{0.050} [.041--.066] & 0.069 [.062--.076]\\
st/6 & $3N/2$ & 0.057$^{1}$ & 0.042$^{1}$ & 0.041$^{1}$ & 0.033$^{1}$\\
st/6 & $2N$ & \textbf{0.014} [.012--.017] & 0.030 [.021--.036] & 0.026 [.021--.030] & 0.021 [.014--.024]\\
st/8 & $N$ & 0.269 [.253--.281] & \textbf{0.236} [.224--.258] & 0.277 [.233--.301] & 0.237 [.217--.269]\\
st/8 & $2N$ & \textbf{0.092} [.088--.099] & 0.109 [.095--.120] & 0.143 [.119--.171] & 0.149 [.116--.163]\\
eo/6 & $N/2$ & 0.101$^{1}$ & 0.051$^{1}$ & 0.099$^{1}$ & 0.064$^{1}$\\
eo/6 & $N$ & \textbf{0.012} [.011--.013] & 0.016 [.014--.022] & 0.026 [.022--.036] & 0.015 [.015--.024]\\
eo/6 & $2N$ & \textbf{0.004}$^{1}$ & 0.005$^{1}$ & 0.012$^{1}$ & \textbf{0.004}$^{1}$\\
\end{tabular}
\end{ruledtabular}
\end{table*}

Below $R=N$, and at depths that are not multiples of $N$, the array or a matching can be ahead: the array at $R=N/2$ for \texttt{eo} ($0.051$ against $0.101$) and the matchings at $R=3N/2$ for \texttt{st}. The reason is visible in the initialisations: the $\pi$-initialised chain is a permutation that maps logical blocks onto logical blocks only at $R=N,2N,\dots$, whereas the shorter diameter of the array and the direct couplings of the matchings help before every pair has met. At $R=2N$, however, the chain is level with or better than every alternative, and for the two-spin chain the margin grows with size: for eight two-spin qubits at $R=2N$ the chain's median is $0.092$ (range $0.088$--$0.099$ over five seeds) against $0.109$ for the array, $0.143$ for the round robin and $0.149$ for random matchings, with the chain's range overlapping only the lower edge of the array's, and for the three-spin chain it is also ahead at $R=N$ ($0.012$ against $0.015$--$0.026$, three seeds) and level at $R=2N$ ($0.004$, single runs, against $0.004$ for the random matching and $0.005$ for the array). For the two-spin chain at $R=N$ it is not: the round robin ($0.050$ against $0.058$ at $n=6$) and the array and the random matchings ($0.236$--$0.237$ against $0.269$ at $n=8$) are ahead by $14$--$16\%$ in the median, with overlapping ranges. The all-to-all coupling that a linear chain provides is therefore not a compromise at $R=2N$: the depth bound holds in any geometry, and at $R=2N$, in the cases tested, the chain optimises at least as well as a two-dimensional array or a fully non-local coupling, while at the minimal depth it is close to, but not always ahead of, them. This is our justification for the chain and for the bubble-sort schedule on it.

\subsection{Other teachers}
\label{sec:other}
To see how far the two prompts generalise we repeated the $n=6$ distillation at a fixed budget (\texttt{st} at $R=2N$, \texttt{eo} at $R=N$, 400/300 steps, one run each) on four further prompts chosen before seeing any result (Table~\ref{tab:other}). The entropies range from $3.96$ to $5.25$ bit and two of the teachers put $15$--$19\%$ of their mass on the all-singlet outcome. The selection rule predicts which initial state has the lower floor, and the KL ordering follows it in every case. For ``Once upon a'' the N\'eel state itself has a floor of $0.20$ in TV---its sector weights $c_0^2=0.143$ and $c_0^2+c_1^2=0.464$ are too small for a teacher with $p_t(|b|=0)=0.186$ and $p_t(|b|\le1)=0.667$---and the two-spin chain stops at TV $0.27$ (KL $0.26$) while the three-spin chain, which has no floor, reaches TV $0.12$ (KL $0.075$). For ``The sky above the city was'' the alternating state has the smaller floor and the smaller KL. For the remaining two prompts the N\'eel two-spin chain reaches or passes the $\chi=4$ level, as for \texttt{cat} (the ordering of alt and N\'eel can depend on the metric: for ``The river flows'' alt has the smaller TV, $0.058$ against $0.065$, and the much larger KL). The three-spin chain at $R=N$ lies between the $\chi=2$ and $\chi=4$ levels for three teachers and below $\chi=4$ for the fourth. These are single runs, subject to the seed-to-seed spread of $\pm20$--$40\%$ seen in Table~\ref{tab:connect}, so only the ordering is claimed. The conclusions of points (a)--(c) of Sec.~\ref{sec:distil} therefore carry over qualitatively, with the quantitative caveat that the floor of a product initial state, N\'eel included, is a property of the teacher's Hamming-class masses and must be checked from Eq.~(\ref{eq:tvfloor}) before the encoding is chosen.

\begin{table*}[t]
\caption{Four additional GPT-2 teachers ($n=6$, 64 outcomes). $H$: entropy (bit); $p_0$: teacher mass on the all-singlet outcome; floor: Eq.~(\ref{eq:tvfloor}) in TV for the alternating ($\ket{ST_0}^{\otimes3}$) and N\'eel initial states; KL (nat) after one run of \texttt{st} ($R=24$, alt from random and N\'eel from $\pi$ initialisation) and \texttt{eo} ($R=18$, $\pi$); MPS reference at $\chi=2/3/4$.}
\label{tab:other}
\begin{ruledtabular}
\begin{tabular}{lcccccccc}
prompt & $H$ & $p_0$ & floor alt & floor N\'eel & st alt & st N\'eel & eo & MPS $\chi{=}2/3/4$\\
\hline
``Once upon a'' ($V{=}8,L{=}2$) & 3.96 & 0.186 & 0.268 & 0.203 & 4.95 & 0.256 & 0.075 & 0.177/0.053/0.029\\
``The sky above the city was'' ($4,3$) & 4.10 & 0.002 & 0.002 & 0.010 & 0.149 & 0.158 & 0.139 & 0.179/0.054/0.020\\
``The river flows'' ($8,2$) & 4.94 & 0.008 & 0.008 & 0.001 & 0.192 & 0.027 & 0.041 & 0.252/0.104/0.030\\
``She opened the door and'' ($4,3$) & 5.25 & 0.153 & 0.153 & 0.010 & 3.81 & 0.014 & 0.008 & 0.199/0.058/0.035\\
\end{tabular}
\end{ruledtabular}
\end{table*}

\subsection{Scaling to eight qubits}
\label{sec:scale}
To see how the picture changes with size we repeated the distillation with eight logical qubits and 256-outcome teachers (\texttt{cat}: top-16 tokens $\times$ 2, entropy $4.45$ bit; \texttt{begin}: top-4 $\times$ 4, $4.63$ bit; MPS references $\chi=1/2/4/8$: $0.56/0.23/0.036/0.005$ and $1.28/0.23/0.025/0.0006$). The two-spin chain ($N=16$) is simulated as before; the three-spin chain ($N=24$, $2^{24}$ amplitudes) is simulated in its $S_z$ sector, $\binom{24}{8}=735\,471$ amplitudes, on which every pulse is the map $\psi\mapsto e^{i\theta/4}[\cos\tfrac\theta2\,\psi-i\sin\tfrac\theta2\,\psi\circ\pi_{ij}]$ with $\pi_{ij}$ the bit-swap permutation of the basis; the readout and the leakage are computed in the same basis (checked against the full simulator at $n=2,3$ to $10^{-16}$).

\begin{figure}[t]
\includegraphics[width=\columnwidth]{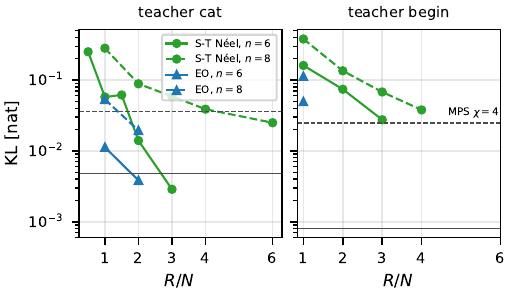}
\caption{KL divergence versus depth in units of the chain length, $R/N$, for six and eight logical qubits (400 Adam steps, 300 for the three-spin chain at $n=6$; N\'eel initial state for the two-spin chain, $\ket{0_L}^{\otimes n}$ for the three-spin chain). Horizontal lines: $\chi=4$ MPS for $n=6$ (solid) and $n=8$ (dashed).}
\label{fig:scale}
\end{figure}

The selection rule again decides the two-spin case, but with a twist: the alternating state $\ket{ST_0}^{\otimes4}$ now has an even number of triplet pairs, so its total spin is even, $S\in\{0,2,4\}$ with weights $0.20,0.57,0.23$, and it is the one-triplet class $|b|=1$ (eight outcomes, mass $0.048$ for \texttt{cat}) that becomes unreachable rather than the all-singlet outcome; the KL stalls at $1.28$. The N\'eel state has weights $0.11,0.27,0.28,0.20,0.10,0.03,0.01,0.001,0$ over $S=0,\dots,8$ and a floor of $10^{-4}$, and with it the two-spin chain again improves geometrically with depth (Fig.~\ref{fig:scale}): $0.28,\,0.088,\,0.059,\,0.039,\,0.025$ at $R=N,2N,3N,4N,6N$ for \texttt{cat} and $0.38,\,0.135,\,0.068,\,0.038$ at $R=N$--$4N$ for \texttt{begin}. The $\chi=4$ level ($0.036$ / $0.025$) is approached at $R=4N$ ($0.039$ / $0.038$, 480 pulses on 16 dots) and, for \texttt{cat}, passed at $R=6N$ ($0.025$), instead of $R\le3N$ at $n=6$; the required depth in units of $N$ grows slowly with $n$. Training length and depth are interchangeable where there is no floor: 2000 instead of 400 steps at $R=3N$ give $0.038$, the same as 400 steps at $R=4N$. For \texttt{begin}, whose one-triplet mass is $0.0004$, the alternating state is again the better start ($0.081$ against $0.135$ at $R=2N$). The three-spin chain from $\ket{0_L}^{\otimes8}$ at $R=N=24$ (276 pulses on 24 dots) reaches $0.053$ for \texttt{cat} ($0.076$ from random initialisation) and $0.113$ for \texttt{begin}, with leakage $0.93$--$0.99$: between the $\chi=2$ and $\chi=4$ levels, whereas at $n=6$ the same depth was within a factor of three of $\chi=4$. At $R=2N=48$ (552 pulses) it reaches $0.019$ (TV $0.039$, leakage $0.97$), below the $\chi=4$ level and below the two-spin chain at $R=6N$ ($0.025$, 720 pulses). Per pulse the two encodings are therefore again level---the two-spin chain needs 360 pulses ($R=3N$) for $0.059$ and 720 for $0.025$---and per dot the two-spin chain is ahead, while the three-spin chain keeps its depth advantage: $\chi=4$ at $R=2N$ against $R=4N$.

\subsection{Ten to fourteen logical qubits on a GPU}
\label{sec:e10}
\begin{figure*}[t]
\includegraphics[width=\textwidth]{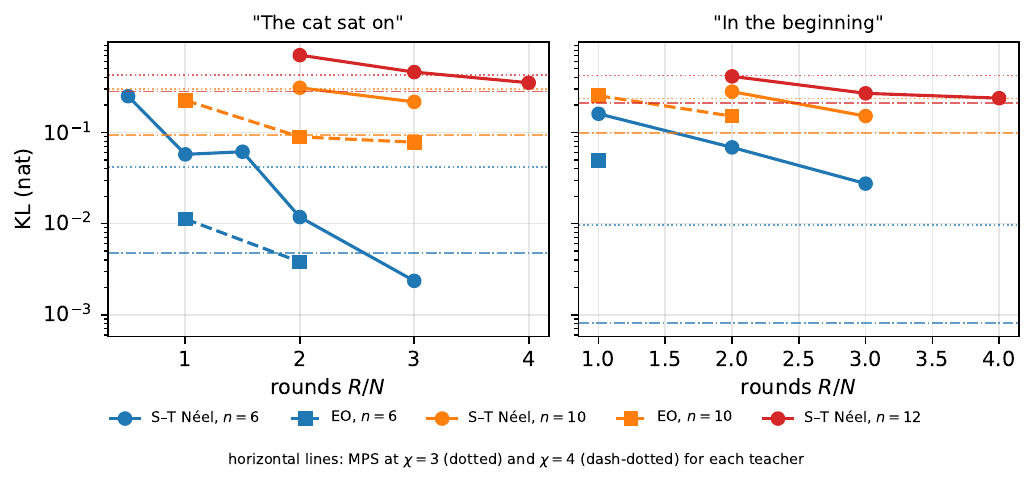}
\caption{KL versus rounds $R/N$ for six, ten and twelve logical qubits (best over initialisations and seeds; two-spin N\'eel chain, solid; three-spin chain, dashed), with the $\chi=3$ (dotted) and $\chi=4$ (dash-dotted) MPS levels of each teacher.}
\label{fig:e10}
\end{figure*}

To see whether the picture survives a larger vocabulary we rebuilt the sector simulator for chains of 20--30 dots: the basis of the fixed-$S_z$ sector is enumerated combinatorially (no $2^N$ array), permutations are looked up by binary search, and the gradient is computed by the adjoint method (two state vectors in memory instead of one per pulse), on a shared GPU with a 4~GB budget (Appendix~\ref{sec:repro}). The teachers are the same two prompts with larger vocabularies: ``The cat sat on'' with $V=32,L=2$ ($n=10$, 1024 outcomes, $4.7$~bit) and $V=64,L=2$ ($n=12$, 4096 outcomes, $4.9$~bit) and $V=128,L=2$ ($n=14$, 16384 outcomes, $5.6$~bit, 28 dots), and ``In the beginning'' with $V=4,L=5$ ($n=10$, $5.8$~bit) and $V=8,L=4$ ($n=12$, $7.2$~bit). Their selection-rule floors from the N\'eel and alternating states are below $0.004$ in TV (the forbidden classes carry little mass for a flat distribution), so these runs test depth and optimisation, not the symmetry. Table~\ref{tab:e10} and Fig.~\ref{fig:e10} summarise the results.

\begin{table*}[t]
\caption{Distillation at $n=6$, $10$, $12$ and $14$ logical qubits: KL (nat, $\mathrm{KL}_\epsilon$) of the two-spin chain from the N\'eel state (400 Adam steps; median [range] over seeds where available---five at $n=6$, three for the two-spin chain at $n=10$ and $12$, two for the three-spin chain at $n=10$---single runs otherwise) and of the three-spin chain from $\ket{0_L}^{\otimes n}$ (300 steps), from the better of the $\pi$ and random initialisations at $n=6$ and from the $\pi$ initialisation at $n\ge10$, against the $\chi=3$ and $\chi=4$ MPS fits of each teacher. Teacher labels give $V,L$. $^\dagger$At $n=12$ the seed-0 run continued to 1000 steps: $0.460\to0.435$ (\texttt{cat}) and $0.269\to0.257$ (\texttt{begin}) at $3N$; $0.352\to0.322$ and $0.240\to0.215$ at $4N$.}
\label{tab:e10}
\footnotesize
\begin{ruledtabular}
\begin{tabular}{l@{\hspace{4pt}}l@{\hspace{4pt}}c@{\hspace{4pt}}c@{\hspace{4pt}}c@{\hspace{4pt}}c@{\hspace{4pt}}c@{\hspace{4pt}}c@{\hspace{4pt}}c}
$n$ & teacher & MPS $\chi{=}3/4$ & st $2N$ & st $3N$ & st $4N$ & eo $N$ & eo $2N$ & eo $3N$\\
\hline
6 & cat $8,2$ & 0.042/0.005 & 0.014 [.012--.017] & 0.004 [.002--.008] & -- & 0.011 & 0.004 & --\\
6 & begin $4,3$ & 0.010/0.001 & 0.082 [.069--.106] & 0.027 & -- & 0.049 & -- & --\\
10 & cat $32,2$ & 0.300/0.095 & 0.335 [.310--.374] & 0.223 [.217--.240] & -- & 0.225 & \textbf{0.092} [.090--.094] & \textbf{0.082} [.078--.085]\\
10 & begin $4,5$ & 0.236/0.099 & 0.288 [.281--.306] & 0.162 [.151--.163] & -- & 0.253 & 0.175 [.151--.199] & --\\
12 & cat $64,2$ & 0.425/0.282 & 0.706 & 0.490 [.460--.574]$^\dagger$ & 0.443 [.352--.477]$^\dagger$ & -- & -- & --\\
12 & begin $8,4$ & 0.421/0.211 & 0.413 & 0.332 [.269--.347]$^\dagger$ & 0.240 [.238--.243]$^\dagger$ & -- & -- & --\\
14 & cat $128,2$ & 0.791/0.512 & 0.956 & 0.700 & 0.592 & -- & -- & --\\
\end{tabular}
\end{ruledtabular}
\end{table*}

Three things change with the vocabulary and one does not. (i) At fixed $R/N$ and fixed training budget the level reached falls with size. For \texttt{cat}, the two-spin chain at $R=3N$ is below $\chi=4$ at $n=6$, between $\chi=3$ and $4$ at $n=10$, and at $\chi=3$ at $n=12$ ($0.49$ against $0.43$), and only at $R=4N$ does the $n=12$ chain pass $\chi=3$ (seed 0: $0.35$; three seeds: $0.35$, $0.44$, $0.48$ against $0.43$, so only one seed in three passes it and the median stays at the $\chi=3$ level); at $n=14$ the chain (28 dots) sits between $\chi=2$ and $3$ at $R=2N$ ($0.96$ against $1.51$ and $0.79$, 756 pulses), passes $\chi=3$ at $R=3N$ ($0.70$, 1134 pulses) and reaches $0.59$ at $R=4N$ (1512 pulses), still above $\chi=4$ ($0.51$). For \texttt{begin} the same chain at $R=3N$ is between $\chi=2$ and $3$ at $n=6$ (this teacher's $\chi=3$ fit is unusually good, $0.010$), between $\chi=3$ and $4$ at $n=10$ and at $n=12$, and reaches $0.24$ (three seeds $0.238$--$0.243$) against the $\chi=4$ value $0.21$ at $R=4N$ ($0.215$ after 1000 steps, i.e.\ the $\chi=4$ level). (ii) In the seed-0 runs at $n=12$ the residual responds more to depth than to the optimisation budget: continuing the $R=3N$ runs from 400 to 1000 steps lowers the KL by $4$--$6\%$ ($0.460\to0.435$, $0.269\to0.257$), whereas one more unit of $R/N$ lowers the same runs by $11\%$ (\texttt{begin}) and $24\%$ (\texttt{cat}); at $R=4N$ the longer training gains $8.5$--$10\%$ ($0.352\to0.322$, $0.240\to0.215$). (iii) The seed-to-seed spread grows from $\pm10\%$ at $n=10$ to up to $\pm20\%$ at $n=12$. What does not change is the depth at which the three-spin chain reaches the $\chi=4$ level: at $R=2N$ (870 pulses, depth 60) it gives $\mathrm{KL}=0.092$ (median of two seeds, $0.090$ and $0.094$) against the $\chi=4$ value $0.095$ for \texttt{cat}, about $3\%$ below it, as it was at or below that level at $R=2N$ for $n=6$, and at $R=3N$ (1305 pulses) it is below it, $0.078$--$0.085$ (two seeds). For \texttt{begin} the three-spin chain gives $0.253$ at $R=N$ and $0.151$--$0.199$ (two seeds) at $R=2N$, between $\chi=3$ and $4$ and comparable to the two-spin chain at $R=3N$, so whether $R=2N$ suffices for the $\chi=4$ level depends on the teacher (for \texttt{cat} both seeds at $R=2N$ are below $0.095$). At $R=N$ (435 pulses on 30 dots) it matches the two-spin chain at $R=3N$ (570 pulses on 20 dots) to within $1\%$ ($0.225$ against $0.223$); at $n=6$ the corresponding pair differed by a factor of three in favour of the two-spin chain. The initial-state ordering also persists where the floors are small: from the N\'eel state the two-spin chain is better than from the alternating state for \texttt{cat} ($0.22$ against $0.37$ at $R=3N$) and equal for \texttt{begin} ($0.16$ against $0.14$), and the two floors that are not small behave as Eq.~(\ref{eq:tvfloor}) says: from $\ket{T_0}^{\otimes10}$ (floor $0.24$ in TV for \texttt{cat}) the chain stops at $\mathrm{TV}=0.53$, and from $\ket S^{\otimes10}$ (floor $0.017$, the one-triplet class being forbidden) it reaches $\mathrm{TV}=0.078$ but $\mathrm{KL}_\epsilon=0.41$, the smoothed divergence being dominated by the structurally zero outcomes. A floor that is small is not sufficient, however: from $\ket{T_0}^{\otimes10}$ the \texttt{begin} teacher (floor $0.004$) stops at $\mathrm{TV}=0.39$ against $0.17$ from the N\'eel state, a finite-depth or optimisation effect rather than a symmetry one. On the GPU a two-spin run takes $2$--$18$ minutes up to $n=12$ and $0.7$--$5.8$ hours at $n=14$, and the three-spin $n=10$ runs $1$--$3.4$ hours at $2.8$~GB, so the 30-dot, 870-pulse network is well within simulation reach; what grows is the depth, and with it the sensitivity to pulse-area noise of Table~\ref{tab:noise}.

\section{Discussion}
\label{sec:disc}

\emph{Two-spin versus three-spin.} The three-spin EO encoding wins on depth and on everything that requires the code space: no symmetry floor, state preparation to $10^{-8}$ at $R=N+2$, unitary compilation to $1-10^{-6}$ at $R=4N$ for the two-qubit blocks tested (Appendix~\ref{sec:unitary}), and distillation to within a factor of three of the $\chi=4$ level already at $R=N$ for six qubits (\texttt{cat}). For the two-spin encoding no fixed initial state reaches every readout distribution (a deterministic all-singlet outcome needs $c_0^2=1$, a deterministic one-triplet outcome $c_1^2=1$); a product of singlets and $T_0$'s forbids whole outcome classes, whereas the N\'eel state forbids none and is limited only by the sector capacities, and with it the encoding reaches $\mathrm{KL}=0.014$ at $R=2N$ (132 pulses, 12 dots) and $0.003$--$0.004$ at $R=3N$ (198 pulses), against $0.011$ at $R=N$ (153 pulses, 18 dots) and $0.004$ at $R=2N$ (306 pulses) for the three-spin encoding; for the second teacher, whose Hamming-class masses happen to suit the alternating state, the two-spin chain trained to convergence reaches $0.007$ at $R=2N$. Per pulse the two encodings are equal and per dot the two-spin chain is ahead; the three-spin advantage is one of depth ($R=N$ instead of $2N$--$3N$) and, for the first teacher, of not having to think about the initial state (for the second the alternating two-spin state is competitive), and for a Born machine that never needs to return to a code space, ``encoding'' reduces to a choice of initial product state and of which pairs are read out.

\emph{Selection rule as a design tool.} Equation~(\ref{eq:rule}) is checked by a state-vector calculation but requires none: from the sector weights of the initial state and the Hamming-class masses of the target one reads off which outcomes are unreachable and by how much, for any depth. It applies to any SU(2)-symmetric processor with singlet/triplet readout, including networks of exchange-coupled dots in two dimensions.

\emph{Hardware.} Six-dot EO devices with exchange-only control and PSB readout exist \cite{Weinstein2023,Intel2025}, a 54-dot array with up to 18 EO qubits has been operated \cite{HRL2026}, and four qubits of the singlet--triplet family ($S$--$T_-$) with baseband exchange control and PSB readout have been demonstrated in germanium \cite{Zhang2025} (see Ref.~\cite{Burkard2023} for a review); the 18-dot chain used here is a $3\times$ extension of the six-dot geometry using the same type of operations (18 exchange layers, PSB on six pairs), and the N\'eel initialisation of the two-spin chain is the one used in exchange-chain experiments \cite{Qiao2020}. We model pulse-area errors as independent perturbations of every $\theta$, a simplification (charge noise is correlated in time and the exchange is nonlinear in the barrier voltage). Adding $\delta_k\sim\mathcal N(0,\sigma^2)$ to every trained pulse area (20 samples, Table~\ref{tab:noise}) changes the KL by $\le0.002$ at $\sigma=0.01$~rad for the six-qubit networks; the increase grows with $\sigma^2$ and with the number of pulses ($+0.004$ at 66 pulses and $+0.010$ at 198 pulses for $\sigma=0.03$~rad, i.e.\ $1\%$ of a $\pi$ pulse). At ten and twelve qubits the same holds with larger increments: $+0.004$ to $+0.013$ at $\sigma=0.01$~rad for 435--1104 pulses, and a factor $3$--$6.5$ at $\sigma=0.1$~rad (Table~\ref{tab:noise}). The consequence is that the optimal depth under noise depends on $\sigma$: at $\sigma=0.03$~rad the best $R=36$ models ($0.003$--$0.004$ noiseless) sit at $0.013$--$0.016$, within a factor of two of the $R=18$--$24$ models ($0.022$--$0.024$), and at $\sigma=0.1$~rad they are worse ($0.17$--$0.21$) than the $R=12$--$18$ ones ($0.09$--$0.11$). The accuracy bought by depth beyond $R\approx2N$ is lost between $\sigma=0.03$ and $0.1$~rad ($1$--$3\%$ of a $\pi$ pulse), which sets the calibration target for a device within this independent-noise model. Only pulse-area noise is modelled here; dephasing, charge noise during the pulses, and readout errors are not, and training the model to be robust to such noise is a design problem in its own right \cite{Ishiyama2022}. Because the network is trained on a simulator and the pulse areas are then fixed, calibration errors can also be absorbed by a short on-device fine-tuning of the same objective. On a device the KL and its gradient would have to be estimated from samples; the exact probabilities and reverse-mode gradients used here are simulator resources, and the shot cost of estimating rare outcomes is not evaluated \cite{Liu2018,Zhu2019} (resolving an outcome of probability $10^{-3}$ to $10\%$ needs of order $10^5$ shots per evaluation). For readout errors, a per-pair PSB error $\varepsilon$ adds up to $n\varepsilon$ in TV, i.e.\ $6\%$ at $n=6$ for $\varepsilon=1\%$, more than the differences between depths reported here, although a known readout error can be folded into the training objective.

\begin{table*}[t]
\caption{KL divergence (nat) after adding independent Gaussian errors of standard deviation $\sigma$ (rad) to every trained pulse area, mean of 20 samples. $p$ = number of pulses. At $\sigma=0.3$~rad the six-qubit networks are at $0.6$--$2.0$. Lower block: ten and twelve logical qubits (Sec.~\ref{sec:e10}; 20 samples, 10 for \texttt{eo}). KL means $\mathrm{KL}_\epsilon$ throughout the tables.}
\label{tab:noise}
\begin{ruledtabular}
\begin{tabular}{llrcccc}
teacher & network & $p$ & $\sigma{=}0$ & 0.01 & 0.03 & 0.1\\
\hline
cat & st N\'eel $R{=}12$ & 66 & 0.058 & 0.058 & 0.062 & 0.114\\
cat & st N\'eel $R{=}24$ & 132 & 0.014 & 0.015 & 0.022 & 0.108\\
cat & st N\'eel $R{=}36$ & 198 & 0.0029 & 0.0038 & 0.0125 & 0.166\\
cat & eo stair $L{=}9$ & 153 & 0.018 & 0.018 & 0.024 & 0.091\\
cat & eo $R{=}36$ & 306 & 0.0038 & 0.0047 & 0.016 & 0.211\\
begin & st alt $R{=}24$\footnote{2000 training steps.} & 132 & 0.0068 & 0.0078 & 0.016 & 0.136\\
begin & st N\'eel $R{=}36$ & 198 & 0.0275 & 0.029 & 0.045 & 0.233\\
begin & eo $R{=}18$ & 153 & 0.049 & 0.051 & 0.060 & 0.169\\
\hline
cat $V{=}32$ ($n{=}10$) & st N\'eel $R{=}60$ & 570 & 0.217 & 0.223 & 0.274 & 0.948\\
begin $V{=}4,L{=}5$ ($n{=}10$) & st N\'eel $R{=}60$ & 570 & 0.151 & 0.157 & 0.199 & 0.762\\
cat $V{=}32$ ($n{=}10$) & eo $R{=}30$ & 435 & 0.225 & 0.229 & 0.256 & 0.699\\
cat $V{=}64$ ($n{=}12$) & st N\'eel $R{=}96$ & 1104 & 0.352 & 0.365 & 0.473 & 1.86\\
begin $V{=}8,L{=}4$ ($n{=}12$) & st N\'eel $R{=}96$ & 1104 & 0.240 & 0.251 & 0.339 & 1.54\\
\end{tabular}
\end{ruledtabular}
\end{table*}

\emph{Scaling.} The $S_z$ conservation of the network means that simulation can be restricted to a single sector, $\binom{18}{6}=18564$ instead of $2^{18}$ amplitudes at $n=6$ and $\binom{24}{8}=735471$ instead of $2^{24}$ at $n=8$; Sec.~\ref{sec:scale} used this to reach 24 dots on a CPU, and Sec.~\ref{sec:e10} 30 dots (ten three-spin qubits, $\binom{30}{10}\approx3\times10^7$ amplitudes) on a shared GPU with a 4~GB budget, by replacing reverse-mode differentiation with the adjoint method so that only two state vectors are held. Beyond that, the network itself is a matrix-product operator of bond dimension at most $4$ per round (each pulse is $a\mathbb 1+b\,\mathrm{SWAP}$, of operator Schmidt rank at most $4$, and one pulse crosses each cut per round), so the low-depth regime $R\lesssim2N$ that matters for hardware may be open to a matrix-product-state simulation whose bond dimension is bounded by $4^{O(R)}$ independently of $N$; whether the actual growth is mild enough to be useful remains to be tested.

\emph{Relation to the SQIF routing.} The bubble-sort schedule appears in the sublinear-resource factoring (SQIF) circuit of Ref.~\cite{Yan2022} purely as routing, with the physics in the ZZ interaction and the SWAPs a cost. On an exchange chain the routing and the interaction are the same pulse, and the exact-cover property becomes a statement about parameters: $R=N$ rounds give every pair of spins one parameter. Whether $R=N$ is enough depends on the target and the encoding, and Tables~\ref{tab:reach} and \ref{tab:distil} give the first answers. Section~\ref{sec:connect} adds the converse: the chain is not what limits the network, since the $N-1$-round bound is geometry-independent and at depth $2N$ nothing we tried optimises better.

\section{Conclusion}
A parallel bubble-sort SWAP network, transplanted from superconducting routing to an exchange-coupled quantum-dot chain, is a one-parameter-per-pulse variational circuit whose depth equals its number of rounds. That depth is within one round of a bound that no arrangement of pairwise exchange pulses can undercut, and at depth $2N$, in the cases tested, the linear chain distils as well as or better than two-dimensional or non-local layouts (at depth $N$, within $20\%$ of them). Its reach is governed by total-spin conservation, which we turned into an explicit and numerically tight selection rule for spin-blockade readout. With that rule, the three-spin exchange-only encoding is shown to distil a GPT-2 next-token distribution to within a factor of three of the bond-dimension-4 matrix-product-state level with 153 pulses on 18 dots, and the two-spin encoding to do as well per pulse, and better per dot, provided the chain starts in the N\'eel product state and is given two to three times the depth (from products of singlets and $T_0$'s it stalls at $0.28$--$1.5$ for the first teacher). The final states lie largely in the leakage space; whether the path through it is necessary was not tested. As a side result, the polar factor of a two-qubit classical weight block can be embedded into the same network (Appendix~\ref{sec:unitary}), while three-qubit blocks already call for approximate compilation. Sector-restricted simulation (Secs.~\ref{sec:scale} and \ref{sec:e10}) carries the same picture to 24--30-dot chains and teachers with up to 16384 outcomes: the three-spin chain reaches the $\chi=4$ level at $R=2N$ at six and, for the first teacher, at ten qubits, while the two-spin chain at fixed $R/N$ loses ground to the MPS as the size grows (from the $\chi=4$ level at six qubits to the $\chi=3$ level at twelve for the first teacher), and at twelve qubits its residual responds more to depth than to the training budget in the runs we could afford. The natural next steps are on-device fine-tuning against pulse-area noise, tensor-network simulation of the low-depth regime beyond 30 dots, and the extension from one distilled distribution to a sequence of layers.

\begin{acknowledgments}
The author thanks George Monma for a detailed review of the manuscript, which led to the sufficiency theorem of Sec.~\ref{sec:suff}, the seed statistics of Table~\ref{tab:connect}, the additional teachers of Sec.~\ref{sec:other} and the definitions of Appendix~\ref{sec:repro}. Numerical experiments were carried out with the assistance of an AI coding agent (Claude, Anthropic) under the author's direction; the manuscript was additionally checked against two further language models (DeepSeek-V4-Flash and OpenAI Codex) whose objections were verified by the author before any change was made. The author is responsible for all claims.
\end{acknowledgments}

\section*{Code and data availability}
The simulation code, the teacher distributions, the trained pulse areas of every run and the raw results behind every table and figure are available at \url{https://github.com/minatoyuichiro/eoall-check} (release \texttt{v1.3}).

\appendix
\section{Compiling logical unitaries and the polar split}
\label{sec:unitary}
The distillation of Sec.~\ref{sec:distil} targets distributions. If instead one wants to embed a \emph{layer} of a classical model, the relevant object is a linear map. A real weight block $W$ has a polar decomposition $W=QP$ with $Q$ orthogonal and $P=\sqrt{W^\top W}$ positive; $Q$ is a unitary that can be compiled into the exchange network, and $P$ (which is diagonal in the right singular basis) would be kept classical, applied to the input before the quantum map (or, as $QPQ^\top$, to the output after it); how the signed linear map $Wx$ is then realised on a register, with its input encoding and normalisation, is not addressed here. For pretrained GPT-2 weights the two factors are far from trivial---$\|P-\mathbb 1\|_F/\|P\|_F\approx0.8$ for a $64\times64$ block of the first attention output projection, and the orthogonal factor has an operator Schmidt spectrum across a balanced bipartition that is, in every block we tested from $4\times4$ to $256\times256$, as flat as that of a Haar-random orthogonal matrix (largest weight $0.015$ at $256\times256$ for both), so $Q$ cannot be shortened by Schmidt truncation either---so, if the split is to be used at all, it would have to be a training-time parametrisation rather than a post-hoc approximation; we do not test such training here and only ask what the compilation of $Q$ costs. This is also the point at which we depart from the tensor-network route of Aizpurua \emph{et al.} \cite{Aizpurua2024,Aizpurua2026}, who write an LLM projection layer as a matrix-product operator and disentangle it with variational unitaries on superconducting hardware, and who reject the polar decomposition because the positive factor of a full layer has large bond dimension: we apply it to small blocks whose positive factor stays classical, in the spirit of orthogonal-layer circuits \cite{Kerenidis2021}; the same split---unitary factor on the circuit, positive factor outside it---appears in the quasi-polar decomposition of quantum neural networks \cite{Ho2026} and in Hamming-weight-preserving orthogonal layers \cite{Monbroussou2025}. We compile single blocks only; how block-wise polar factors are composed into a full layer is left open.

We compiled $Q$ into the three-spin network for $n=2$ ($N=6$, $d=4$): the objective is the process fidelity $|\tr(Q^\dagger M)|^2/d^2$ of the logical block $M=B^\dagger\mathcal B_R B$ ($B$ the $2^N\times2^n$ code isometry), which is reduced by leakage as well as by coherent logical error, since $M$ is unitary only when the map is leakage-free. Targets are the polar factors of three Gaussian random $4\times4$ matrices and of the $4\times4$ corner of the GPT-2 block. Results (best of 3 L-BFGS restarts): $R=8$: $0.62$--$0.90$; $R=12$: $0.88$--$0.96$; $R=16$: $0.97$--$0.99$; $R=20$: $0.997$--$0.9997$; $R=24$ (60 pulses): $F_{\rm proc}=1-10^{-6}$, leakage $10^{-13}$, for all four targets. A leakage-free compilation of these four real orthogonal targets (six parameters each in $O(4)$; complex $SU(4)$ targets were not tested) on two three-spin qubits thus took $R=4N$ rounds (60 pulses in 24 layers, against 22 pulses in 13 layers for the exact CNOT of Fong and Wandzura \cite{Fong2011}, 19--28 pulses for topology-optimised CX sequences \cite{Chadwick2025}, and the analytic controlled-rotation families of Ref.~\cite{Zeuch2020}, all of which target one gate rather than an arbitrary $SU(4)$ element), three times the $R=N+2$ needed for a state, consistent with the count of parameters ($6$ for $O(4)$, $15$ for $SU(4)$, versus $6$ for a state) and with the need to control the action on the whole code space rather than on one vector. For $n=3$ ($N=9$, $d=8$, 63 real parameters in $SU(8)$) the same procedure does not reach an exact compilation within the depths we could afford: the four targets give $F_{\rm proc}=0.40$--$0.50$ at $R=18$ (72 pulses), $0.55$--$0.68$ at $R=27$ (108), $0.59$--$0.85$ at $R=36$ (144), $0.66$--$0.86$ at $R=45$ (180) and $0.72$--$0.73$ at $R=54$ (216), with leakage $0.1$--$0.35$ (Fig.~\ref{fig:reach}). To separate expressivity from optimisation we repeated the $n=3$ compilation with six restarts of Adam followed by L-BFGS, in the $S_z=\tfrac32$ sector that contains the code space ($\binom{9}{3}=84$ amplitudes). The Haar and GPT-2 factors then give $F_{\rm proc}=0.65$ at $R=36$ (the $0.85$ above is a single fortunate restart; the six new restarts lie between $0.51$ and $0.65$), $0.73$ at $R=54$ and $0.91$ at $R=90$ (360 pulses), with the six restarts within $0.035$ of each other from $R=54$ on; the relative error $\|\tilde QP-W\|_F/\|W\|_F$ of the reconstructed block ($\tilde Q$: the phase-aligned logical block $M$, not itself unitary) is $0.34$--$0.38$, $0.21$--$0.22$ and $0.07$--$0.08$ (Haar and GPT-2 factors). At $R=36$ the identity and a product $Q_a\otimes Q_b$ of a two-qubit and a one-qubit orthogonal matrix reach $F_{\rm proc}\ge1-2\times10^{-4}$ from every restart (best $1-10^{-12}$), and so do three \emph{planted} targets, obtained by first optimising the network for zero leakage and maximal operator entanglement across the first qubit (largest operator-Schmidt weight $0.25$, the minimum; this measures the operator's entanglement, not its entangling power) and then handing the resulting logical unitary to fresh restarts: every reachable target we gave the optimiser at $R=4N$ was found. We could not reach the Haar and GPT-2 factors at $R=4N$, whereas every planted target was recovered; we take this as evidence, not proof, that they lie outside the leakage-free reachable set at that depth (the planted targets probe one family of reachable unitaries, not the whole set). Beyond $R=4N$ the situation reverses: at $R=6N$ the planting itself stalls at leakage $0.2$, and at $R=10N$ the same optimiser returns $0.91$ even for the identity, which is reachable at $\theta\equiv0$. The numbers at $R>4N$ are thus optimisation-limited, and the depth at which a generic element of $SU(8)$ becomes reachable is not determined by these runs. For the polar split this means that at $n\ge3$ the unitary factor should be compiled approximately, or block by block on two-qubit units, rather than exactly.

\section{Reproducibility}
\label{sec:repro}
\emph{Teachers.} GPT-2 small (\texttt{gpt2}, 124M parameters, Hugging Face \texttt{transformers} 5.16.1 with \texttt{GPT2TokenizerFast}), exact softmax at temperature 1. For a prompt, the vocabulary is the set of the $V$ most probable next tokens after the prompt; it is kept fixed for all $L$ steps, and at every step the GPT-2 next-token distribution conditioned on the prompt and the tokens chosen so far is restricted to that vocabulary and renormalised. The teacher $p_t(t_1,\dots,t_L)$ is the product of these conditionals over all $V^L$ sequences. A sequence is encoded as $L$ blocks of $\log_2V$ bits, each block the rank of the token in the vocabulary (most probable first); the bit string of length $n=L\log_2V$ is read from the chain with token $t_k$ occupying logical qubits $(k-1)\log_2V+1,\dots,k\log_2V$, most significant bit first. Prompts: ``The cat sat on'' ($V=8$, $L=2$), ``In the beginning'' ($V=4$, $L=3$); for $n=8$, $V=16,L=2$ and $V=4,L=4$ with the same prompts; Sec.~\ref{sec:other} adds ``Once upon a'' ($8,2$), ``The sky above the city was'' ($4,3$), ``The river flows'' ($8,2$) and ``She opened the door and'' ($4,3$). The stored teacher tables are normalised to $1$ within $10^{-7}$. The exact teacher tables are stored in the repository.

\emph{Optimisation.} Exchange-network parameters are initialised from \texttt{torch} generator seed 0 (\texttt{random}: uniform in $[0,2\pi)$; \texttt{0} and \texttt{pi}: $\mathcal N(0,0.1^2)$ around $0$ or $\pi$; the layout comparison of Sec.~\ref{sec:connect} uses seeds 0--4 for \texttt{st} and 0--2 for \texttt{eo}, and the seed statistics quoted in Sec.~\ref{sec:distil}(b) use seeds 0--4), Adam with learning rate $0.05$ for 300 (\texttt{eo}) or 400 (\texttt{st}) steps unless stated, loss $\mathrm{KL}_\epsilon$ with $\epsilon=10^{-12}$. MPS Born machines: Adam, 3000 steps, best of seeds 0--2. L-BFGS (state preparation and unitary compilation): \texttt{torch} L-BFGS with strong-Wolfe line search, history 50, up to 500--600 iterations, restarts as stated; the sector-restricted compilations of Appendix~\ref{sec:unitary} use 200 Adam steps followed by 400 L-BFGS iterations from seeds 700--705, and the planted targets are obtained by minimising leakage plus $0.5\times$ the largest operator-Schmidt weight across the first qubit from seeds 9500--9503. The simulations of Secs.~\ref{sec:network}--\ref{sec:scale} are double-precision state-vector (or, for 24 dots, $S_z$-sector) calculations on CPU (\texttt{torch} 2.13.0, \texttt{numpy} 2.2.6, \texttt{scipy} 1.15.3); the GPT-2 teachers are generated at the model's default single precision. Noise entries in Table~\ref{tab:noise} are means over 20 samples (10 for the three-spin $n=10$ row); the per-sample values are stored for the $n\ge10$ rows only. Every KL, TV and leakage quoted for a trained network is evaluated on the stored pulse areas of that network (except the two overwritten $n=12$ endpoints noted below) (an earlier version of the training loop stored the parameters one optimiser step past the recorded minimum; the stored circuits were re-evaluated, which changed quoted values by at most $0.002$ nat).

\emph{Large chains (Sec.~\ref{sec:e10}).} \texttt{e10\_sector.py}: the fixed-$S_z$ basis is enumerated by a vectorised recursion over sites and stored as sorted integers; the permutation of a pulse is found by binary search; the gradient is the adjoint-method recursion $\lambda_{k-1}=U_k^\dagger\lambda_k$ started from $\lambda_K=\partial L/\partial\mathrm{Re}\,\psi_K+i\,\partial L/\partial\mathrm{Im}\,\psi_K$ ($L$ the loss, $K$ the number of pulses, $\psi_K$ the state after the last pulse), $\partial L/\partial\theta_k=\mathrm{Re}\braket{\lambda_k|\partial_{\theta_k}U_k\,\psi_{k-1}}$ with the readout gradient written by hand, verified against automatic differentiation to $10^{-15}$ at $n=6$. Runs at 30 and 28 dots use complex single precision (verified to $10^{-6}$); all others double. Seeds 0--2 as for $n=6$; the $n=12$ 1000-step entries continue the seed-0 400-step run with the same optimiser state (the continuation shared the checkpoint file of the 400-step run; the released script names checkpoints by step count, so a continuation now requires copying the checkpoint, and the stored parameters of those two 400-step endpoints were overwritten, their KL and TV being taken from the run logs). Hardware: one RTX~6000~Ada shared with other users (4~GB budget), \texttt{torch} 2.6.0+cu124.

\emph{Code.} Every table and figure can be regenerated from the scripts listed in the repository's README; the manuscript corresponds to the tagged release \texttt{v1.3} (commit \texttt{94d4f87}) of \url{https://github.com/minatoyuichiro/eoall-check}.

\bibliography{apssamp}

\end{document}